\documentclass[aps,12pt,eqsecnum, preprint,nofootinbib,superscriptaddress]{revtex4}
\usepackage{amssymb,amsmath,amsthm,graphicx,amscd,mathtools}
\usepackage[mathscr]{eucal}
\usepackage{upgreek,enumerate,color,verbatim,comment,ulem,bigints,stackrel}
\usepackage{bm,pbox,cancel}
\usepackage[hidelinks]{hyperref}

\begin{document}

\title{\Large  Quantum Radiation and Dissipation in Relation to \\ Classical Radiation and Radiation Reaction}
\author{Jen-Tsung Hsiang}
\email{cosmology@gmail.com}
\affiliation{Center for High Energy and High Field Physics, National Central University, Taoyuan 320317, Taiwan, ROC}
\author{Bei-Lok Hu}
\email{blhu@umd.edu}
\affiliation{Maryland Center for Fundamental Physics and Joint Quantum Institute,  University of Maryland, College Park, Maryland 20742, USA}

\begin{abstract}
This work continues the investigation of radiation phenomena from atom-field interactions, extending our earlier study~\cite{QRad} of quantum radiation from a stationary atom's internal degree of freedom, modeled by a harmonic oscillator, to the emittance of classical radiation. By assuming that the atom interacts with a quantum scalar field initially in a coherent state, we show how a stochastic component of the internal dynamics of the atom arises from the vacuum fluctuations of the field, resulting in the emittance of quantum radiation, whose reaction induces quantum dissipation in the internal dynamics. We also show how the deterministic mean field drives the internal classical mean component to emit classical radiation and receive classical radiation reaction. Both components are statistically distinct and fully decoupled. It is clearly seen that the effects of the vacuum fluctuations of the field are matched with those of quantum radiation reaction, not with classical radiation reaction, as the folklore goes. In contrast to the quantum component of the atom's internal dynamics, which always equilibrates, the relaxation dynamics of the classical component largely depends on the late-time behavior of the mean field. For the values of the parameters defining the coherent state of the field much greater than unity, if the mean field remains periodic, then the internal dynamics of the atom will appear classical and periodic. If the mean field diminishes with time, then the classical component of the atom's internal dynamics subsides but the quantum component will abide and dynamically equilibrate. This also explains why quantum radiation from a stationary atom is not observed, and a probe located far away only sees classical radiation. Our analysis therefore paints a continuum landscape starting from vacuum fluctuations in the quantum field to classical radiation and radiation reaction.
\end{abstract}
\maketitle

\baselineskip=18pt
\allowdisplaybreaks
\numberwithin{equation}{section}

\section{Introduction}

In an earlier paper~\cite{QRad} (Paper I), we considered several fundamental issues related to the quantum vacuum~\cite{MilBook}  in the context of atom-field interactions~\cite{PassanteBook,CTBook,ScullyBook} pertaining to quantum radiative processes.  We begin with the vacuum fluctuations in a quantum  field, examine how they bring  forth stochastic motion in the internal degrees of freedom (idf) of an atom, induce  it to emit quantum radiation and how the emitting of radiation generates a reactive force on the atom. All this are quantum in nature. {When the atom-field system reaches equilibrium, the energy flow drained by this reactive force in the idf of the atom is balanced by the power fed by vacuum fluctuations of the field. Meanwhile, as shown in Paper I, at places sufficiently far away from the atom, the outgoing energy of quantum radiation is compensated by an incoming energy flux}. A powerful relation which one can use to check the energy flux balance  in the relevant  processes is the fluctuation-dissipation relation (FDR). Under equilibrium conditions, referring to the quantum field alone, vacuum fluctuations are related to quantum dissipation~\cite{RHA,RHK,JH1,CPR2D,CPR4D,NLFDR,FDRSq}  by a fluctuation-dissipation relation\footnote{Note two sets of FDRs are involved. When we address the vacuum fluctuations vs. radiation reaction, we refer to the FDR of the free environment field (in its initial state). However, when we compare the FDR in linear response theory vs. nonequilibrium dynamics, it is the FDR of the system (in its final equilibrium state). A more thorough discussion can be found in~\cite{QTD1}.}.  

\subsection{Basic Issues and Prior Treatments}

\paragraph{Fluctuation Dissipation Relations}

The FDR in the quantum field is easy to show, as has been done in the 90s~\cite{RHA,RHK,JH1}. It amounts to the relation between the Hadamard function and the retarded Green function. A greater challenge is to show whether and how  a FDR relation exists  for the atom in the context of the atom-field interaction, namely, between {its} quantum fluctuations, and quantum dissipation in the dynamics of the idf of the atom~\cite{CPR2D,CPR4D,NLFDR,FDRSq}. This lesser known process goes as follows:  Under specific conditions, as we shall explain in the main text, radiation of a quantum nature is emitted from the atom. The reactive force from this quantum radiation engenders quantum dissipation in the dynamics of the atom's idf, and under equilibrium conditions,  a quantum fluctuation-dissipation relation (FDR) of the atom can be shown to  exist reflecting the  {subtle} balance between the {dissipated energy by the quantum self-force and inputted energy from the surrounding quantum field fluctuations}. 

\paragraph{Some Conventional Misconceptions}

It is often said that vacuum fluctuations and classical radiation reaction are two sides of the same coin, and one can approach it from either side as they are related by a fluctuation-dissipation relation.   In Paper I we   pointed out the flaws in this view.  To begin with,  the former entity is quantum while the latter classical.  One cannot connect them prima face,  a big divide exists between these two levels of theoretical structure. A simple observation is: there are situations where {classical} radiation reaction can be zero, such as in a uniformly accelerated charge, which emits classical radiation~\cite{Jackson}, and yet vacuum fluctuations are always present. There is indeed a relation between vacuum fluctuations and quantum dissipation --- one can call this {\it quantum} radiation reaction,  but not with {\it classical} radiation reaction.  

\paragraph{From Vacuum Fluctuations in the Field to Quantum Dissipation in the Atom} 
These  rather intricate quantum processes are not often discussed, but can be understood in two stages:  a)   fluctuations in a quantum field induces a stochastic component in the dynamics of the idf of an atom. This motion causes the atom to emit quantum radiation. b) the backaction in the emittance of quantum radiation generates a reactive force on the atom, which shows up as  quantum dissipation in the atom's idf.  All this are at the quantum level, not related to classical radiation and radiation reaction.  
The primary purpose of Paper I was to trace out the subtle relations between fluctuations in the quantum field and quantum radiation from the atom, and how that is related to quantum dissipation. 
In fact we considered a stationary atom and show that there is a delicate yet exact energy flux balance  due to the correlation between  quantum radiation  emitted from the atom and the vacuum fluctuations of the quantum field at the position of a detector placed far away from the atom. 

\subsection{This Work: Objective and Procedures}

\paragraph{How does this chain of quantum events lead to classical radiation?}

 In this paper we follow this line of reasoning and set forth to show how  classical radiation and radiation reaction appear from the the quantum field by assuming that the field is in a coherent state.  The energy flux expressions we shall derive have as the leading contribution a classical term. This enables us to bring the above quantum story starting from vacuum fluctuations all the way to  classical radiation while still keeping all the quantum attributes as described in Paper I. With this we can explicitly address all the issues we mentioned above, vacuum fluctuations to quantum dissipation to quantum radiation to classical radiation and classical radiation reaction.  Attempts to connect quantum radiation in the nature of Unruh effect~\cite{Unr76,DeW79}, namely, thermal radiance felt by a uniformly accelerated atom, with classical (Lamor) radiation registered by a detector or probe at some distance from this atom have been made. For example  Landulfo, Fulling and Matsas \cite{Fulling19} found a relation between Unruh's thermal radiance in, and the Larmor radiation emitted by, a uniformly accelerated charge, the latter can be seen as entirely built from zero-Rindler-energy modes, thus bridging the quantum and the classical radiations.  We are not aware of attempts to connect vacuum fluctuations in a quantum field with quantum dissipation in an atom to quantum and classical radiation from an atom.

\paragraph{Main Features in our Findings}

{We find that for a stationary atom  whose internal degree of freedom is modeled by a harmonic oscillator and is coupled to a massless scalar field initially prepared in a coherent state, the dynamics of the field and the atom's internal motion can both be decomposed into the corresponding classical, mean component and the fluctuating, quantum component. Both components of the atom's internal degree of freedom act like a driven damped oscillator. The classical component is driven by the mean field but damped by a classical  reactive force from the emission of classical radiation. In contrast, the quantum component is steered by the vacuum fluctuations of the field. Its motion engenders quantum radiation, and in return a quantum dissipation is exerted on the quantum component of the internal dynamics.} 

{The classical deterministic motion of the atom's internal degree of freedom in general dominates over the quantum motion at late times if the coherent parameters{, that is, the parameters defining the coherent state of the field, e.g.~$\{\alpha\}$ in \eqref{E:fgkbdrt},}  are mostly much greater than unity, with the exception that the mean free field decays with time. This happens when the mean field is not periodic and has a wide, sufficiently smooth spectrum. In the latter case, the internal dynamics of the atom is quantum mechanical at late times.  These two cases are different not only in their statistical nature, but also in their ability to reach equilibration. Periodic driving due to the classical mean component of the free field will in general induce a periodic internal motion of the atom, so the internal dynamics will not have an equilibrium state. On the other hand   a diminishing mean field can allow the quantum dynamics of the internal degree of freedom to prevail at late times, and equilibrate. The ability of the quantum component inside the atom suggests that the contribution of quantum dissipation, or quantum radiation reaction, is completely made up by the corresponding contribution due to vacuum fluctuations of the free field.}

{The radiation emitted by the atom can likewise be separated into a classical and a quantum component. In general the classical radiation component {either} will not settle down, or it just plainly slackens off, depends on the properties of the mean field at late times. Conversely, at late times the energy flux from the quantum component of the emitted radiation measured at a distance sufficiently far away from the stationary atom will be balanced out by another incoming energy flow. This was a key discovery we found in Paper I and~\cite{LH06}:  The source of this incoming flux results from the correlation in the quantum components between the distant radiation field, that is, the far-field component, and the local free field around the atom. The amount of correlation to make this happen is enforced by the fluctuation-dissipation relations associated with the quantum components of the field and the internal dynamics of the atom. Such a balance in the energy exchange between the atom and the field is generically not available for the classical mean components because they are deterministic, there are no fluctuations,  and thus no FDRs.  Therefore, {in short, quantum radiation is always present, but for a stationary atom its effect at spatial infinity is completely cancelled out, so it cannot be observed.} The probe located far away from a stationary atom thus only sees classical radiation.}  More detailed summary can be found in the last section.

\paragraph{Organization}
{In Sec.~\ref{S:lskfmsmfjb}, while considering the dynamics of the atom's internal degrees of freedom coupled to a scalar field in the coherent state, we carefully separate the classical mean component and the quantum fluctuating component in the field and the internal dynamics. We  highlight their differences {in terms of their deterministic versus stochastic natures}, and discuss their implications in approaching dynamical equilibration. In Sec.~\ref{S:rithvgf}, we study the late-time behavior of the internal dynamics of its classical and quantum components, and the condition that may lead to stationarity in its correlation function. Then we examine in Sec.~\ref{S:bgvruts} how a late-time equilibrium state can or cannot be reached from the aspect of the reduced dynamics of the atom's internal motion driven by the quantum field in its coherent state. In Sec.~\ref{S:eobgd}, we switch over to the perspective of the field, and investigate the radiation power that flows to spatial infinity. In passing we reassert the significant role the FDRs plays in bringing the quantum components of radiation power from a stationary atom to balance,  and the consequence of their absence in the classical component. Finally,  in conclusion, we highlight the similarity and disparity between the classical mean and the quantum fluctuating dynamics in the atom-field system we have considered, and discuss their implications in late-time dynamics.}




\section{internal dynamics of the atom with backaction from a quantum field}\label{S:lskfmsmfjb}

We consider a harmonic  atom whose internal degree of freedom (idf) $Q(t)$ is modeled by a quantum harmonic oscillator of mass $m$ and bare frequency $\omega_{\textsc{b}}$.  Let $Q(t)$  be coupled to a massless scalar field $\phi(x)$ in Minkowski spacetime with $x=(\bm{x},t)$.  The total action of this atom-field interacting system takes the form
\begin{align}\label{E:kdfhgbdw}
	S&=\int\!d^{4}x\sqrt{-\eta}\,\Bigl[-\frac{1}{2}\,\eta^{\mu\nu}\partial_{\mu}\phi(x)\,\partial_{\nu}\phi(x)\Bigr]+\int\!d^{4}x\sqrt{-\eta}\;J(x)\phi(x)\notag\\
	&\qquad\qquad\qquad\qquad\qquad\qquad\qquad\qquad+\int\!dt\;\Bigl[\frac{m}{2}\dot{Q}^{2}(t)-\frac{m\omega_{\textsc{b}}^{2}}{2}Q^{2}(t)\Bigr]\,,
\end{align}
with the scalar current $J(x^{\mu})=e\,Q(t)\delta^{(3)}(\bm{x}-\bm{z})$, the coupling strength $e$, the spatial position of the harmonic atom $\bm{z}$, and the metric tensor $\eta^{\mu\nu}=\operatorname{diag}(-1,+1,+1,+1)$. We suppose for the moment that the external degree of freedom $z^{\mu}$ of the atom is non-dynamical but prescribed. When we promote the canonical variables in the action \eqref{E:kdfhgbdw} to quantum operators, we arrive at a simultaneous set of Heisenberg equations
\begin{align}
	\square\hat{\phi}(x)&=-e\,\hat{Q}(t)\delta^{(3)}(\bm{x}-\bm{z})\,,\label{E:lgndlg3}\\
	\ddot{\hat{Q}}(t)+\omega_{\textsc{b}}^{2}\hat{Q}(t)&=\frac{e}{m}\,\hat{\phi}(\bm{z},t)\,.\label{E:lgndlg4}
\end{align}
Since they are linear equations, the corresponding operator solutions to \eqref{E:lgndlg3} can readily be found: 
\begin{align}\label{E:gnbskfgjs}
	\hat{\phi}(x)=\hat{\phi}_{\text{h}}(x)+\int\!d^{4}x'\;G_{0,\textsc{R}}^{(\phi)}(x,x')\,\hat{J}(x')\,,
\end{align}
where $\hat{\phi}_{\text{h}}(x)$ is the homogeneous, in-field solution, satisfying the free-field wave equation $\square\hat{\phi}_{\text{h}}(x)=0$, and $G_{0,\textsc{R}}^{(\phi)}(x,x')$ is the retarded Green's function of the free field, obeying the inhomogeneous wave equation 
\begin{equation}
	\square G_{0,\textsc{R}}^{(\phi)}(x,x')=-\frac{1}{(-\eta)^{\frac{1}{4}}}\,\delta^{(4)}(x-x')\,\frac{1}{(-\eta')^{\frac{1}{4}}}\,.
\end{equation}
It is important to mark the difference between $\hat{\phi}(x)$ and $\hat{\phi}_{\text{h}}(x)$, and their roles in the atom-field dynamics. The former contains an additional contribution of the radiation field due to the internal dynamics of the atom, conveyed by the second term in \eqref{E:gnbskfgjs}.

Eq.~\eqref{E:gnbskfgjs} allows us to rewrite \eqref{E:lgndlg4} into
\begin{align}
	\ddot{\hat{Q}}(t)+\omega_{\textsc{b}}^{2}\hat{Q}(t)-\frac{e^{2}}{m}\int_{0}^{t}\!dt'\;G_{0,\textsc{R}}^{(\phi)}(\bm{z},t,\bm{z},t')\,\hat{Q}(t')&=\frac{e}{m}\,\hat{\phi}_{\text{h}}(\bm{z},t)\,.\label{E:bgweirusd}
\end{align}
Its solution assumes the general form
\begin{align}\label{E:gkdfgjsd}
	\hat{Q}(t)&=\hat{Q}_{\text{h}}(t)+\hat{Q}_{\text{inh}}(t)\,,&&\text{with}&\hat{Q}_{\text{inh}}(t)&=e\int_{0}^{t}\!dt'\;G_{\textsc{R}}^{(Q)}(t-t')\,\hat{\phi}_{\text{h}}(\bm{z},t')\,,
\end{align}
where $\hat{Q}_{\text{h}}(t)=d_{1}(t)\,\hat{Q}(0)+d_{2}(t)\,\dot{\hat{Q}}(0)$ is the homogeneous solution to \eqref{E:bgweirusd} and depends on the initial conditions of $\hat{Q}$. Two fundamental solutions $d_{1}(t)$ and $d_{2}(t)$ are tailored for the initial-value problem of \eqref{E:bgweirusd}, so they are chosen to take on special values at the initial times $t=0$ according to $d_{1}(0)=1$, $\dot{d}_{1}(0)=0$, $d_{2}(0)=0$, and $\dot{d}_{2}(0)=1$. An overhead dot represents taking the derivative with respect to $t$.

The inhomogeneous (particular) solution $\hat{Q}_{\text{inh}}(t)$ is caused by the free quantum field $\hat{\phi}_{\text{h}}$. It rises from zero once the atom-field interaction is turned on.  Since it does not depend on the initial conditions of $\hat{Q}(t)$  it has a statistics different from the homogeneous part $\hat{Q}_{\text{h}}(t)$.

\subsection{Retarded Green Function}

The retarded Green's function $G_{\textsc{R}}^{(Q)}(\tau)=d_{2}(\tau)/m$ of the internal degree of freedom $\hat{Q}$, on account of the interaction with the field, satisfies
\begin{equation}\label{E:dgksdhser}
	\frac{d^{2}}{dt^{2}}G_{\textsc{R}}^{(Q)}(t-s)+\omega_{\textsc{b}}^{2}G_{\textsc{R}}^{(Q)}(t-s)-\frac{e^{2}}{m}\int_{0}^{t}\!dt'\;G_{0,\textsc{R}}^{(\phi)}(\bm{z},t;\bm{z},t')\,G_{\textsc{R}}^{(Q)}(t'-s)=\delta(t-s)\,.
\end{equation}
When $\hat{\phi}$ is a massless scalar field, we may reduce the equation of motion \eqref{E:bgweirusd} to a local form. Since
\begin{align}
	G_{0,\textsc{R}}^{(\phi)}(\bm{z},t;\bm{z},t')=i\,\theta(t-t')\bigl[\hat{\phi}_{\text{h}}(\bm{z},t),\hat{\phi}_{\text{h}}(\bm{z},t')\bigr]=-\frac{1}{2\pi}\,\theta(\tau)\,\delta'(\tau)\,,
\end{align}
the backaction on $\hat{Q}$ due to the inhomogeneous part of $\hat{\phi}$ is given by
\begin{align}\label{E:poijdkjf1}
	-e^{2}\int_{0}^{t}\!ds\;G_{0,\textsc{R}}^{(\phi)}(\bm{z},t;\bm{z},s)\,\hat{Q}(s)&=\frac{e^{2}}{2\pi}\biggl\{-\delta(0)\,\hat{Q}(t)+\delta(t)\,\hat{Q}(0)+\int_{0}^{t}\!ds\;\delta(t-s)\,\dot{Q}(s)\biggr\}\,.
\end{align}
We note that the integral
\begin{equation}
	\frac{e^{2}}{2\pi}\int_{0}^{t}\!ds\;\delta(t-s)\,\dot{Q}(s)=\frac{e^{2}}{4\pi}\,\dot{Q}(t)=2m\gamma\,\dot{Q}(t)
\end{equation}
gives the frictional force. 

On the other hand, the general solution to \eqref{E:lgndlg3} can also be expressed in term of the out-field and the advanced Green's function
\begin{equation}
	\hat{\phi}(\bm{z},t)=\hat{\phi}_{\textsc{out}}(\bm{z},t)+e\int_{0}^{t}\!ds\;G^{(\phi)}_{0,\textsc{A}}(\bm{z},t;\bm{z},s)\,\hat{Q}(s)\,,
\end{equation}
where $G^{(\phi)}_{0,\textsc{A}}(\bm{x},t;\bm{x}',t')=G^{(\phi)}_{0,\textsc{A}}(\bm{x}',t';\bm{x},t)$. In the limit $\bm{x}'\to \bm{x}\to\bm{z}$, we find
\begin{equation}
	G_{0,\textsc{A}}^{(\phi)}(\bm{z},t;\bm{z},t')=\frac{1}{2\pi}\,\theta(-\tau)\,\delta'(\tau)\,,
\end{equation}
so the corresponding backaction comes from
\begin{align}\label{E:poijdkjf2}
	-e^{2}\int_{t}^{\infty}\!ds\;G_{0,\textsc{A}}^{(\phi)}(\bm{z},t;\bm{z},s)\,\hat{Q}(s)&=\frac{e^{2}}{2\pi}\biggl\{\delta(-\infty)\,\hat{Q}(\infty)-\delta(0)\,\hat{Q}(t)-\int_{t}^{\infty}\!ds\;\delta(t-s)\,\dot{Q}(s)\biggr\}\,.
\end{align}
For times not at the asymptotic past and future, we may write \eqref{E:poijdkjf1} and \eqref{E:poijdkjf2} as
\begin{align}
	-e^{2}\int_{0}^{t}\!ds\;G_{0,\textsc{R}}^{(\phi)}(\bm{z},t;\bm{z},s)\,\hat{Q}(s)&=-\frac{e^{2}}{2\pi}\,\delta(0)\,\hat{Q}(t)+2m\gamma\,\dot{Q}(t)\,,\\
	-e^{2}\int_{t}^{\infty}\!ds\;G_{0,\textsc{A}}^{(\phi)}(\bm{z},t;\bm{z},s)\,\hat{Q}(s)&=-\frac{e^{2}}{2\pi}\,\delta(0)\,\hat{Q}(t)-2m\gamma\,\dot{Q}(t)\,,
\end{align}
and find that the frictional force can be given by the combination
\begin{equation}
	2m\gamma\,\dot{Q}(t)=-e^{2}\int_{0}^{t}\!ds\;\frac{1}{2}\Bigl[G_{0,\textsc{R}}^{(\phi)}(\bm{z},t;\bm{z},s)-G_{0,\textsc{A}}^{(\phi)}(\bm{z},t;\bm{z},s)\Bigr]\,\hat{Q}(s)\,.
\end{equation}
This provides an alternative perspective why this mysterious expression has been used to calculate the self-force on the atom due to the   radiation~\cite{Rohrlich}. 

\subsection{Radiation Reaction}

Back to \eqref{E:poijdkjf1}, the term proportional to $\delta(0)$ will be absorbed into $\omega_{\textsc{b}}^{2}$ to form the physical frequency
\begin{equation}
	\omega_{\textsc{r}}=\omega_{\textsc{b}}-\frac{e^{2}}{2\pi}\,\delta(0)\,,
\end{equation}
such that \eqref{E:bgweirusd} becomes
\begin{equation}\label{E:lmdius}
	\ddot{\hat{Q}}(t)+2\gamma\,\dot{\hat{Q}}(t)+\omega_{\textsc{r}}^{2}\hat{Q}(t)=\frac{e}{m}\,\hat{\phi}_{\text{h}}(\bm{z},t)\,.
\end{equation}
This form is convenient to work with, and in particular,  the Fourier transform of $G_{\textsc{R}}^{(Q)}(\tau)$ has a rather simple form
\begin{equation}\label{E:gbjdfhd}
	\tilde{G}_{\textsc{R}}^{(Q)}(\kappa)=\int_{-\infty}^{\infty}\!d\tau\;G_{\textsc{R}}^{(Q)}(\tau)\,e^{i\kappa\tau}=\frac{1}{m(-\kappa^{2}+\omega^{2}_{\textsc{r}}-i\,2\gamma\kappa)}\,.
\end{equation}
From the derivation of \eqref{E:lmdius}, we see that the damping emerges as the reaction force to the radiation emitted by the atom. Together with the fluctuating force $e\,\hat{\phi}_{\text{h}}(\bm{z},t)$, they are consequence of the atom-field interaction. Thus, the damping term and the driving term will account for the energy exchange between the atom and the field. The force $e\,\hat{\phi}_{\text{h}}(\bm{z},t)$ associated with the free field pumps energy to the atom, while $-2m\gamma\,\dot{\hat{Q}}(t)$ dissipates the atom's energy in the form of radiation field. This relation is what we shall focus on when  examining the time variation of the energy exchange between the atom and the surrounding field to see it comes to an exact balance at late times.
	
\subsection{Quantum field initially in a coherent state}

An important feature of $\hat{Q}_{\text{h}}(t)$ is that it exponentially decays with time over the relaxation time scale $\gamma^{-1}$ with $\gamma=e^{2}/8\pi m$, due to the damping in \eqref{E:lmdius}.  At late times $t\gg\gamma^{-1}$, the second term in \eqref{E:gkdfgjsd} survives and thus the late-time dynamics of $\hat{Q}$ is exclusively governed by the free scalar field $\hat{\phi}_{\text{h}}$. This has important implications which we shall see later.

In the Heisenberg picture, we consider the case where  the initial state of the total system at $t=0$ is a direct product state made up of an arbitrary, normalized oscillator state $\hat{\rho}^{(Q)}$ and the multi-mode coherent state\footnote{The essential properties of the coherent state are reviewed in Appendix~\ref{S:rtghjvdf}.} of the field $\lvert\{\alpha\}\rangle$,
\begin{equation}\label{E:fgkbdrt}
	\hat{\rho}(0)=\hat{\rho}^{(Q)}(0)\otimes\lvert\{\alpha\}\rangle\langle\{\alpha\}\rvert\,,
\end{equation}
where $\lvert\{\alpha\}\rangle$ is the shorthand notation for the multi-mode coherent state, $\lvert\{\alpha\}\rangle=\lvert\alpha_{\bm{k}_{1}}\rangle\otimes\lvert\alpha_{\bm{k}_{2}}\rangle\otimes\cdots$ where $\alpha_{\bm{k}_{i}}\in\mathbb{C}$ and $\bm{k}_{i}$   labels a particular field mode. The expectation value $\langle\cdots\rangle=\operatorname{Tr}\{\hat{\rho}(0)\cdots\}$ will be defined with respect to this initial state. The coherent state has distinct features among the Gaussian states of the field. The field in this state has a nonzero expectation value but  minimal field fluctuations, equal to its vacuum fluctuations. Hence the coherent state is sometimes referred to as the `most classical' quantum state and often treated so when issues of  quantum-to-classical correspondence are discussed. However,  it is important to remember that the nonequilibrium system under consideration is inherently quantum,  even if the field is assumed to be in a coherent state, because once the interaction begins, the state of the whole system will become highly entangled, a feature totally absent in the corresponding classical interacting systems.

Assume that $\hat{\phi}_{\text{h}}$ has a plane-wave expansion
\begin{equation}\label{E:skgbsfer}
	\hat{\phi}_{\text{h}}(\bm{x},t)=\int\!\frac{d^{3}\bm{k}}{(2\pi)^{\frac{3}{2}}}\;\frac{1}{\sqrt{2\omega}}\,\Bigl[\hat{a}_{\bm{k}}^{\vphantom{\dagger}}e^{+i\bm{k}\cdot\bm{x}}e^{-i\omega t}+\hat{a}_{\bm{k}}^{\dagger}e^{-i\bm{k}\cdot\bm{x}}e^{+i\omega t}\Bigr]\,,
\end{equation}
where $\omega=\lvert\bm{k}\rvert$, and $\hat{a}_{\bm{k}}^{\vphantom{\dagger}}$, $\hat{a}_{\bm{k}}^{\dagger}$ are the annihilation and creation operators of each mode $\bm{k}$. Its expectation value is given by
\begin{equation}
	\varphi_{\text{h}}(x)=\operatorname{Tr}\Bigl\{\hat{\rho}(0)\,\hat{\phi}_{\text{h}}(x)\Bigr\}=\langle\hat{\phi}_{\text{h}}(x)\rangle=\int\!\frac{d^{3}\bm{k}}{(2\pi)^{\frac{3}{2}}}\;\frac{1}{\sqrt{2\omega}}\,\bigl(\alpha_{\bm{k}}^{\vphantom{\dagger}}e^{+i\bm{k}\cdot\bm{x}-i\omega t}+\alpha_{\bm{k}}^{*}e^{-i\bm{k}\cdot\bm{x}+i\omega t}\bigr)\,,\label{E:kgeusds}
\end{equation}
with $\alpha_{\bm{k}}^{\vphantom{\dagger}}=\langle\{\alpha\}\vert\hat{a}_{\bm{k}}^{\vphantom{\dagger}}\vert\{\alpha\}\rangle$. This looks like a classical, free scalar field in plane-wave expansion, and the amplitude of each mode is proportional to $\alpha_{\bm{k}}$. In turn, taking the same expectation value of \eqref{E:lmdius} gives an equation of motion
\begin{equation}\label{E:lmdius2}
	\ddot{\mathsf{Q}}(t)+2\gamma\,\dot{\mathsf{Q}}(t)+\omega_{\textsc{r}}^{2}\mathsf{Q}(t)=\frac{e}{m}\,\varphi_{\text{h}}(\bm{z},t)\,,
\end{equation}
with $\mathsf{Q}(t)=\operatorname{Tr}\bigl\{\hat{\rho}(0)\,\hat{Q}(t)\bigr\}$. The mean dynamics of $\hat{Q}(t)$ is mathematically identical to the dynamics of a damped classical harmonic oscillator, driven by a deterministic force $e\,\varphi_{\text{h}}(\bm{z},t)$, evaluated at the location of the atom. The deviation from the mean dynamics is then described by $\hat{q}=\hat{Q}-\mathsf{Q}$, following a similar equation of motion
\begin{equation}\label{E:lmdius3}
	\ddot{\hat{q}}(t)+2\gamma\,\dot{\hat{q}}(t)+\omega_{\textsc{r}}^{2}\hat{q}(t)=\frac{e}{m}\,\bigl[\hat{\phi}_{\text{h}}(\bm{z},t)-\varphi_{\text{h}}(\bm{z},t)\bigr]\,.
\end{equation}
obtained by subtracting \eqref{E:lmdius2} from \eqref{E:lmdius}. The operator $\hat{q}$ has a vanishing expectation value $\langle\hat{q}(t)\rangle=0$, but as shown later, it describes the quantum fluctuations of the internal dynamics.

The rendition above may lead one to think that when $\lvert\alpha_{\bm{k}}\rvert\gg1$, apart from minute zero-point fluctuations, the reduced mean dynamics of $\hat{Q}$ driven by the quantum field in the coherent state is more or less equivalent to the reduced dynamics of a classical field $\varphi_{\text{h}}(\bm{z},t)$. This is not entirely correct, especially when   quantum field fluctuation effects are involved. The first hint comes from the two-point function of the field.

The two-point function of $\hat{\phi}_{\text{h}}(x)$ is given by
\begin{align}
	\operatorname{Tr}\Bigl\{\hat{\rho}(0)\,\hat{\phi}_{\text{h}}(x)\hat{\phi}_{\text{h}}(x')\Bigr\}&=\varphi_{\text{h}}(x)\varphi_{\text{h}}(x')+\langle0\vert\,\hat{\phi}_{\text{h}}(x)\hat{\phi}_{\text{h}}(x')\,\vert0\rangle\,.\label{E:fkghbsjrd}
\end{align}
The second term on the righthand side \eqref{E:fkghbsjrd} is the corresponding two-point function of $\hat{\phi}_{\text{h}}(x)$ in the vacuum state. On the other hand, the form of $\varphi_{\text{h}}$ in \eqref{E:kgeusds} may suggest identifying the first term as the two-point function of $\varphi_{\text{h}}(x)$. This is problematic conceptually. Since $\varphi_{\text{h}}(x)$ is a deterministic $c$-number field, it contains no random elements and hence it should not have the two-point correlation function. 

\noindent { \bf Deterministic vs Stochastic Variables}

This  may be a good place to clarify the differences between  deterministic and stochastic (random) variables. Suppose $F(t)$ is a deterministic variable of the physical quantity we are interested in. We make a measurement of $F$ at time $t$, and another measurement at $t'$. If we repeat the same set of measurements on another identically prepared copy, and another again, we always get the same outcomes at times $t$ and $t'$. On the other hand, if $\mathcal{F}(t)$ is a stochastic variable, then repeated measurements will give different sets of outcomes at $t$ and $t'$ for each realization. The correlation function of a stochastic variable then tells us how the result at $t'$ is qualitatively correlated with the one at $t$, or to infer what we might obtain at time $t'$, given the measurement result at time $t$.  For a deterministic variable, there is no such uncertainty.

One may argue that since $\varphi_{\text{h}}(x)$ contains fast oscillating modes it inevitably introduces uncertainty, and one may define its correlation function by an appropriate time average
\begin{equation}
	\langle\varphi_{\text{h}}(x)\varphi_{\text{h}}(x')\rangle_{\textsc{t}}=\lim_{T\to\infty}\frac{1}{2T}\int_{-T}^{T}\!ds\;\varphi(\bm{x}',t+s)\varphi(\bm{x},t'+s)\,.
\end{equation}
Plugging in \eqref{E:kgeusds} gives
\begin{align}
	\langle\varphi(x)\varphi(x')\rangle_{\textsc{t}}=\sum_{\bm{k}}\bigg\{\alpha_{\bm{k}}^{\vphantom{*}}\alpha_{\bm{k}}^{*}e^{+i\bm{k}\cdot(\bm{x}-\bm{x}')}e^{-i\omega(t-t')}+\alpha_{\bm{k}}^{\vphantom{*}}\alpha_{-\bm{k}}^{*}e^{+i\bm{k}\cdot(\bm{x}+\bm{x}')}e^{-i\omega(t-t')}+\text{C.C.}\biggr\}\,.
\end{align}
where we also have used
\begin{equation}
	\lim_{T\to\infty}\frac{1}{2T}\int_{-T}^{T}\!ds\;e^{i(\omega-\omega')s}=\lim_{T\to\infty}\frac{\sin(\omega-\omega')T}{(\omega-\omega')T}=\delta_{\omega,\omega'}\,.
\end{equation}
Here we assume the mode label $\bm{k}$ to be discrete for convenience. The continuum limit of the mode sum is understood as
\begin{equation}
	\sum_{\bm{k}}=\int\!\frac{d^{3}\bm{k}}{(2\pi)^{\frac{3}{2}}}\;\frac{1}{\sqrt{2\omega}}\,.
\end{equation}
Since both $\omega=\lvert\bm{k}\rvert$ and $\omega'=\lvert\bm{k}'\rvert$ are positive, the terms proportional to the product $\alpha_{\bm{k}}\alpha_{\bm{k}'}$ and its complex conjugate vanish. The two-point function obtained by taking the time average turns out to be always stationary in time, which ostensibly is  different from the first term in \eqref{E:fkghbsjrd}. Thus this argument fails.

Furthermore, for any $c$-number function $\psi(x)$, one can always compute
\begin{equation}
	\operatorname{Tr}\Bigl\{\hat{\rho}(0)\,\psi(x)\psi(x')\Bigr\}=\psi(x)\psi(x')
\end{equation}
in the same fashion as \eqref{E:fkghbsjrd}. However, we would not regard this as a correlation function of $\psi$ per se, in the sense of the above discussions. These observations suggest  that  a coherent state description and the classical formulation are not exactly equivalent.  We shall return to this point later. For now let us proceed to study the late-time behavior of the internal dynamics of the harmonic atom when it is coupled to a quantum field initially in a coherent state.

\subsection{Internal dynamics of the harmonic atom at late-times}  

Complementary to \eqref{E:gkdfgjsd}, the general solutions to \eqref{E:lmdius2} and \eqref{E:lmdius3} take the forms
\begin{align}
	\mathsf{Q}(t)&=\mathsf{Q}_{\text{h}}(t)+e\int_{0}^{t}\!dt'\;G_{\textsc{R}}^{(Q)}(t-t')\,\varphi_{\text{h}}(\bm{z},t')\,,\label{E:iefjdn}\\
	\hat{q}(t)&=\hat{q}_{\text{h}}(t)+e\int_{0}^{t}\!dt'\;G_{\textsc{R}}^{(Q)}(t-t')\,\bigl[\hat{\phi}_{\text{h}}(\bm{z},t)-\varphi_{\text{h}}(\bm{z},t)\bigr]\,.
\end{align}
The homogeneous solutions $\mathsf{Q}_{\text{h}}(t)$, $\hat{q}_{\text{h}}(t)$ have similar exponentially decaying behavior with increasing time, as $\hat{Q}_{\text{h}}(t)$ does. Then the two-point function $\langle\hat{Q}(t)\hat{Q}(t)\rangle$ of the internal dynamics can be written as
\begin{align}\label{E:oerjfhdo}
	\langle\hat{Q}(t)\hat{Q}(t')\rangle=\mathsf{Q}(t)\mathsf{Q}(t')+\langle\hat{q}(t)\hat{q}(t')\rangle\,.
\end{align}
In particular when $t'\to t$, we find the second moment of $\hat{Q}$ has the form $\langle\hat{Q}^{2}(t)\rangle=\mathsf{Q}^{2}(t)+\langle\hat{q}^{2}(t)\rangle$ and thus $\langle\hat{q}^{2}(t)\rangle$ gives the dispersion of $\hat{Q}$, and $\hat{q}(t)$ accounts for the non-deterministic component of $\hat{Q}$. If we use \eqref{E:gkdfgjsd} to express $\hat{Q}(t)$ as $\hat{Q}_{\text{h}}(t)+\hat{Q}_{\text{inh}}(t)$, then the two-point function \eqref{E:oerjfhdo} can be alternatively written as
\begin{align}
	\langle\hat{Q}(t)\hat{Q}(t')\rangle=\langle\hat{Q}_{\text{h}}(t)\hat{Q}_{\text{h}}(t')\rangle+\langle\hat{Q}_{\text{inh}}(t)\hat{Q}_{\text{h}}(t')\rangle+\langle\hat{Q}_{\text{h}}(t)\hat{Q}_{\text{inh}}(t')\rangle+\langle\hat{Q}_{\text{inh}}(t)\hat{Q}_{\text{inh}}(t')\rangle\,.
\end{align}
Observe that $\hat{Q}_{\text{h}}(t)$ and $\hat{Q}_{\text{inh}}(t')$ commute since $\hat{Q}(0)$ and $\hat{\phi}_{\text{h}}(x)$ are assumed to be statistically independent. We then have, say, $\langle\hat{Q}_{\text{h}}(t)\hat{Q}_{\text{inh}}(t')\rangle=\mathsf{Q}_{\text{h}}(t)\mathsf{Q}_{\text{inh}}(t)$. It vanishes at late times, as well as $\langle\hat{Q}_{\text{h}}(t)\hat{Q}_{\text{h}}(t')\rangle$, so in this limit the remaining nonvanishing component of the two-point function \eqref{E:oerjfhdo} is
\begin{equation}
	\langle\hat{Q}(t)\hat{Q}(t')\rangle\simeq\langle\hat{Q}_{\text{inh}}(t)\hat{Q}_{\text{inh}}(t')\rangle\,,\label{E:gkfjh}
\end{equation}
for $t$, $t'\to\infty$.

\section{late-time stationarity of the reduced dynamics of the atom's idf}\label{S:rithvgf}

The behaviors of the Green's functions of $\hat{Q}$ play an important role when we try to determine whether equilibration conditions of the $\hat{Q}$ dynamics exist.  We already have worked out $G_{\textsc{R}}^{(Q)}(t-t')$, so now we focus on $G_{\textsc{H}}^{(Q)}(t,t')$. We will show that in general $G_{\textsc{H}}^{(Q)}(t,t')$ is not translationally-invariant in time due to the nonequilibrium dynamics of $\hat{Q}$, but at sufficiently late times, it can gradually become  a stationary function of time, that is, $G_{\textsc{H}}^{(Q)}(t,t')\to G_{\textsc{H}}^{(Q)}(t-t')$, under certain conditions.

\subsection{Behavior of the Hadamard function}

The Hadamard function $G_{\textsc{H}}^{(Q)}(t,t')$ of $\hat{Q}$ can be readily written down with the help of Eq.~\eqref{E:gkdfgjsd},
\begin{align}
	G_{\textsc{H}}^{(Q)}(t,t')&=\frac{1}{2}\langle\bigl\{\hat{Q}(t),\hat{Q}(t')\bigr\}\rangle\notag\\
	&=\cdots+e^{2}\int_{0}^{t}\!ds\int_{0}^{t'}\!ds'\;G_{\textsc{R}}^{(Q)}(t-s)\,G_{\textsc{R}}^{(Q)}(t'-s')\,\prescript{\alpha}{}{G}_{0,\textsc{H}}^{(\phi)}(\bm{z},s;\bm{z},s')\,.\label{E:rtufgbv}
\end{align}
The $\dots$ represents terms that have negligible contributions at late times. Using \eqref{E:fkghbsjrd}, we can in general write the Hadamard function $\prescript{\alpha}{}{G}_{0,\textsc{H}}^{(\phi)}(\bm{z},s;\bm{z}',s')$ of the free scalar field $\hat{\phi}_{\text{h}}$ in the coherent state as,
\begin{align}\label{E:gldjgfkd}
	\prescript{\alpha}{}{G}_{0,\textsc{H}}^{(\phi)}(\bm{z},s;\bm{z}',s')=\varphi_{\text{h}}(\bm{z},s)\varphi_{\text{h}}(\bm{z}',s')+\prescript{\textsc{vac}}{}{G}_{0,\textsc{H}}^{(\phi)}(\bm{z},s;\bm{z}',s')\,.
\end{align}
Here the left superscript denotes the state of the field, and when the field under consideration is a free field we add a $0$ to the right subscript. The other super- and sub-scripts are self-explanatory.

Thus Eq.~\eqref{E:rtufgbv} becomes
\begin{align}
	G_{\textsc{H}}^{(Q)}(t,t')&=\cdots+e^{2}\int_{0}^{t}\!ds\int_{0}^{t'}\!ds'\;G_{\textsc{R}}^{(Q)}(t-s)\,G_{\textsc{R}}^{(Q)}(t'-s')\,\varphi_{\text{h}}(\bm{z},s)\varphi_{\text{h}}(\bm{z}',s')\label{E:rjgfmjnfg}\\
	&\qquad\qquad\qquad+e^{2}\int_{0}^{t}\!ds\int_{0}^{t'}\!ds'\;G_{\textsc{R}}^{(Q)}(t-s)\,G_{\textsc{R}}^{(Q)}(t'-s')\,\prescript{\textsc{vac}}{}{G}_{0,\textsc{H}}^{(\phi)}(\bm{z},s;\bm{z},s')\,.\notag
\end{align}
The first term on the righthand side of \eqref{E:rjgfmjnfg} is nothing but $\mathsf{Q}_{\text{inh}}(t)\mathsf{Q}_{\text{inh}}(t')$, a deterministic component in $G_{\textsc{H}}^{(Q)}(t,t')$. It is worth further investigation because it may turn out to be the main contribution to the nonstationary behavior of $G_{\textsc{H}}^{(Q)}(t,t')$ at late times. Define
\begin{align}\label{E:iurgs}
	f(t;\omega)=m\int_{0}^{t}\!ds\;G_{\textsc{R}}^{(Q)}(t-s)\,e^{-i\omega s}\,.
\end{align}
We obtain
\begin{align}
	\mathsf{Q}_{\text{inh}}(t)=\frac{e^{2}}{m^{2}}\int\!\frac{d^{3}\bm{k}}{(2\pi)^{\frac{3}{2}}}\frac{1}{\sqrt{2\omega}}\;\Bigl[\alpha_{\bm{k}}^{\vphantom{\dagger}}e^{+i\bm{k}\cdot\bm{z}}f(t;\omega)+\alpha_{\bm{k}}^{*}e^{-i\bm{k}\cdot\bm{z}}f^{*}(t;\omega)\Bigr]\,,\label{E:euijfg}
\end{align}
so let us examine the late-time behavior of $\mathsf{Q}_{\text{inh}}(t)$.

By \eqref{E:gbjdfhd}, we find
\begin{align}\label{E:yqeus}
	f(t;\omega)=\tilde{d}_{2}(\omega)\,e^{-i\omega t}\Bigl\{1-e^{i\omega t}\,d_{1}(t)+i\,\omega\,e^{i\omega t}\,d_{2}(t)\Bigr\}\,,
\end{align}
and $f^{*}(t;\omega)=f(t;-\omega)$. For simplicity, we put the origin of the external coordinate system at the location of the atom, i.e., $\bm{z}=0$, and assume that the spacetime is homogeneous and isotropic. Then we can reduce $\mathsf{Q}_{\text{inh}}(t)$ into
\begin{align}
	\mathsf{Q}_{\text{inh}}(t)=\frac{8\sqrt{\pi}\gamma}{m}\int_{0}^{\infty}\!d\omega\;\omega^{\frac{3}{2}}\Bigl[\alpha_{\omega}^{\vphantom{\dagger}}\,f(t;\omega)+\alpha_{\omega}^{*}\,f^{*}(t;\omega)\Bigr]\,,\label{E:dkjbsss}
\end{align}
Here we include $\varphi_{\text{h}}(\bm{x},t)$ in \eqref{E:kgeusds} for comparison
\begin{align}
	\varphi_{\text{h}}(\bm{x},t)&=\frac{1}{\sqrt{\pi}r}\int_{0}^{\infty}\!d\omega\;\omega^{\frac{1}{2}}\Bigl[\alpha_{\omega}^{\vphantom{*}}\,\sin\omega r\,e^{-i\omega t}+\alpha_{\omega}^{*}\,\sin\omega r\,e^{+i\omega t}\Bigr]\,,\label{E:kbgeugfd}
\end{align}
with $r=\lvert\bm{x}\rvert$.

The long-time behavior of $\mathsf{Q}_{\text{inh}}(t)$  depends sensitively on the choice of $\alpha_{\omega}$ in the expansion of $\varphi_{\text{h}}$. When, for example, the coherent parameter $\alpha_{\omega}$ is proportional to $\delta(\omega-\varpi)$, then the driving force in \eqref{E:lmdius2} becomes a single-mode sinusoidal function of time with frequency $\varpi$, then \eqref{E:lmdius2} describes a damped harmonic oscillator driven by a sinusoidal force. It is well known that for $t$ greater than $\gamma^{-1}$, $\mathsf{Q}(t)\simeq\mathsf{Q}_{\text{inh}}(t)$ will oscillate at the frequency $\varpi$ and has an amplitude proportional to $\varpi^{\frac{3}{2}}\tilde{d}_{2}(\varpi)$. Thus the exponentially decaying behavior of $G_{\textsc{R}}^{(Q)}(t)$ is not in general sufficient to damp $\mathsf{Q}_{\text{inh}}(t)$ at large $t$ even though $\varphi_{\text{h}}(\bm{0},t)$ is not an exponentially growing function of $t$.

\subsection{Illustrative examples}

In passing, it is interesting to mention a few more examples before we delve into the general considerations. 

1.  Suppose that $\varphi_{\text{h}}(\bm{0},t)$ is a {\it monochromatic light}
\begin{equation}
	\varphi_{\text{h}}(\bm{0},t)=\frac{\pi}{2}\,\cos\bigl(2\pi\,\frac{t}{\tau}\bigr)\,,
\end{equation}
with the amplitude $A=\pi/2$, period $\tau=2$ or angular frequency $\omega_{\textsc{d}}=\pi$. The inhomogeneous solution of the displacement $\mathsf{Q}(t)$ is given by
\begin{align}
	\mathsf{Q}_{\text{inh}}(t)&\to\frac{2e}{m}\,A\gamma\omega_{\textsc{d}}\,\lvert\tilde{d}_{2}(\omega_{\textsc{d}})\rvert^{2}\,\sin\omega_{\textsc{d}} t\,,\\
\intertext{and}
	\dot{\mathsf{Q}}_{\text{inh}}(t)&\to\frac{2e}{m}\,A\gamma\omega_{\textsc{d}}^{2}\,\lvert\tilde{d}_{2}(\omega_{\textsc{d}})\rvert^{2}\,\cos\omega_{\textsc{d}} t\,,
\end{align}
when $t\gg\gamma^{-1}$. Then the mechanical energy of the driven damped oscillator at late times becomes
\begin{equation}
	E_{\textsc{mech}}=\frac{m}{2}\,\dot{\mathsf{Q}}^{2}_{\text{inh}}(t)+\frac{m\omega_{\textsc{r}}^{2}}{2}\,\mathsf{Q}^{2}_{\text{inh}}(t)=16\pi A^{2}\gamma^{3}\omega_{\textsc{d}}^{2}\lvert\tilde{d}_{2}(\omega_{\textsc{d}})\rvert^{4}\,\Bigl[\omega_{\textsc{d}}^{2}\cos\omega_{\textsc{d}} t+\omega_{\textsc{r}}^{2}\sin\omega_{\textsc{d}} t\Bigr]\,.
\end{equation}
We note that it will be a constant $16\pi A^{2}\gamma^{3}\omega_{\textsc{d}}^{4}\lvert\tilde{d}_{2}(\omega_{\textsc{d}})\rvert^{4}$ at late time when the driving frequency $\omega_{\textsc{d}}$ matches the physical frequency $\omega_{\textsc{r}}$, not the resonance frequency $\Omega=\sqrt{\omega_{\textsc{r}}^{2}-\gamma^{2}}$, where $\tilde{d}_{2}(\omega)$ has a maximum. This does not mean the motion reaches equilibrium, a stronger constraint like the existence of the FDR or stationarity is needed. Rather, it says that at late times, the energy pumped into the oscillator happens to be equal to the dissipated energy. This does not occur for a general periodic motion. 

2.  As a second example,  consider a {\it sawtooth wave}
\begin{equation}\label{E:dfgkbf}
	\varphi_{\text{h}}(\bm{0},t)=i\,\frac{A}{\pi}\Bigl[\ln(1-e^{+i\omega_{\textsc{d}} t})-\ln(1-e^{-i\omega_{\textsc{d}} t})\Bigr]=A-\frac{2A}{\tau}\bigl(\,t\!\!\!\mod \tau\bigr)\,.
\end{equation}
The corresponding $\mathsf{Q}_{\text{inh}}(t)$ is
\begin{align}
	\mathsf{Q}_{\text{inh}}(t)&=\frac{e}{m}\frac{A}{2\pi\Omega\omega_{\textsc{d}}^{3}}\Bigl\{-2e^{-\gamma t}\Bigl[\bigl(\pi\omega_{\textsc{d}}+2\gamma\bigr)\Omega\,\cos\Omega t+\bigl(\pi\gamma\omega_{\textsc{d}}+\gamma^{2}-\Omega^{2}\bigr)\sin\Omega t\Bigr]+4\gamma\Omega\Bigr.\notag\\
	&\qquad\qquad\qquad\qquad\qquad\qquad+\Bigl.i\,2\omega_{\textsc{d}}\Omega\Bigl[\ln(1-e^{+i\omega_{\textsc{d}} t})-\ln(1-e^{-i\omega_{\textsc{d}} t})\Bigr]\Bigr\}\,,
\end{align}
and
\begin{align}
	\dot{\mathsf{Q}}_{\text{inh}}(t)&=\frac{e}{m}\frac{A}{\pi\Omega\omega_{\textsc{d}}}\Bigl\{e^{-\gamma t}\Bigl[\Omega\,\cos\Omega t+\bigl(\gamma+\pi\omega_{\textsc{d}}\bigr)\sin\Omega t\Bigr]-\Omega\Bigr\}\to-\frac{eA}{\pi m\omega_{\textsc{d}}}\,.
\end{align}
At late times $\mathsf{Q}_{\text{inh}}(t)$ also takes on a sawtooth shape, but $\dot{\mathsf{Q}}_{\text{inh}}(t)$ approaches a constant. At   first sight, it seems strange that $\dot{\mathsf{Q}}_{\text{inh}}(t)$ takes on a negative constant value, inconsistent with the periodic, sawtooth-like motion described by $\mathsf{Q}_{\text{inh}}(t)$. In truth, the consistency is preserved because the force $e\varphi_{\text{h}}(\bm{0},t)$ in \eqref{E:dfgkbf} instantaneously brings $\mathsf{Q}_{\text{inh}}(t)$ to $+\frac{eA}{m\omega_{\textsc{d}}^{2}}$ at $t=n\tau^{+}$, and then the oscillator moves at the constant speed $\dot{\mathsf{Q}}_{\text{inh}}(\infty)$ to $-\frac{eA}{m\omega_{\textsc{d}}^{2}}$ at $t=(n+1)\tau^{-}$. Hence the periodic motion is realized this way. Obviously the corresponding mechanical energy cannot be a constant at late times

3.  Next, it would be interesting to examine the case that $\varphi_{\text{h}}(\bm{0},t)$ is a {\it real periodic functionof time} $g(t)$ , satisfying $g(t)=g(t+\tau)$. The parameter $\tau$ is the period. One of the previous examples is a special case,  that  $g(t)$ is a trigonometric function. Let us first look into the frequency spectrum of the periodic function $g(t)$. The Fourier transform $\tilde{g}(\omega)$ is given by
\begin{align}
	\tilde{g}(\omega)=\int_{-\infty}^{\infty}\!ds\;g(t)\,e^{i\omega t}&=\sum_{n=-\infty}^{\infty}\int_{0}^{\tau}\!ds\;g(t+n\tau)\,e^{i\omega (t+n\tau)}=2\pi\int_{0}^{\tau}\!ds\;g(t)\,e^{i\omega t}\sum_{k=-\infty}^{\infty}\delta(\omega\tau-2k\pi)\,,\label{E:fghjfgsd}
\end{align}
where we have assumed that the order of summation and integration is exchangeable, and that the summation
\begin{equation}
	\delta(t)=\frac{1}{2\pi}\,\sum_{n=-\infty}^{\infty}e^{int}\,,
\end{equation}
converges to the delta function. In fact the summation also applies to
\begin{equation}
	\sum_{n=-\infty}^{\infty}e^{int}=\sum_{n=-\infty}^{\infty}e^{in(t-2k\pi)}=2\pi\delta(t-2k\pi)\,,
\end{equation}
for $k\in\mathbb{Z}$ since $e^{i2k\pi}=1$. If we let
\begin{equation}
	\tilde{g}_{\tau}(\omega)=\int_{0}^{\tau}\!dt\;g(t)\,e^{i\omega t}\,,
\end{equation}
then we can write \eqref{E:fghjfgsd} as
\begin{equation}
	\tilde{g}(\omega)=\tilde{g}_{\tau}(\omega)\,\frac{2\pi}{\tau}\sum_{k=-\infty}^{\infty}\delta(\omega-\frac{2k\pi}{\tau})\,,
\end{equation}
with $\tilde{g}_{\tau}^{*}(\omega)=\tilde{g}_{\tau}(-\omega)$. Hence the spectrum is comb-like,  with spikes located at $\omega=k\,\nu$, spacing $\nu=\frac{2\pi}{\tau}$, and has a height proportional to $\tilde{g}_{\tau}(\omega)\,\frac{2\pi}{\tau}$. If we carry out the inverse Fourier transform to find $g(t)$, we obtain
\begin{align}
	g(t)=\int_{-\infty}^{\infty}\!\frac{d\omega}{2\pi}\;\tilde{g}(\omega)\,e^{-i\omega t}&=\sum_{k=-\infty}^{\infty}g_{k}\,e^{-ik\nu t}\,,\label{E:dsjfge}
\end{align}
and 
\begin{align}
	g_{k}&=\frac{1}{\tau}\,\tilde{g}_{\tau}(\frac{2k\pi}{\tau})=\frac{1}{\tau}\int_{0}^{\tau}\!dt\;g(t)\,e^{i\frac{2k\pi}{\tau}t}\,,&&\text{and}&g_{k}^{*}&=g_{-k}^{\vphantom{*}}\,.
\end{align}
Eq.~\eqref{E:dsjfge} is the Fourier series of the function $g(t)$ defined only with the finite time interval $t\in[0,\tau)$, and $g_{k}$ are the corresponding coefficients. Thus it implies that a periodic function can be constructed by shifting this function by $\tau$ an infinite number of times along the real $t$ axis in both directions. It also says that a periodic function of period $\tau$ can always be written as a Fourier series over the fundamental frequency $\nu$ with a suitable choice of the expansion coefficients $g_{k}$. Note that since $g(t)$ is real, we also have $\tilde{g}^{*}(\omega)=\tilde{g}(-\omega)$.

Now suppose $\varphi_{\text{h}}(\bm{0},t)$ is such a periodic function. Then we can write $\mathsf{Q}_{\text{inh}}(t)$ as
\begin{equation}
	\mathsf{Q}_{\text{inh}}(t)=\sum_{k=-\infty}^{\infty}\varphi_{k}\int_{0}^{t}\!ds\;d_{2}(t-s)\,e^{-ik\nu s}=\sum_{k=-\infty}^{\infty}\varphi_{k}\,f(t;k\nu)\,,
\end{equation}
by \eqref{E:iurgs} and \eqref{E:yqeus}, where $\varphi_{k}$ is the corresponding Fourier coefficient of $\varphi_{\text{h}}(\bm{0},t)$. In the limit $t\to\infty$, we have $f(t;k\nu)\simeq\tilde{d}_{2}(k\nu)\,e^{-ik\nu t}$, and hence
\begin{equation}
	\mathsf{Q}_{\text{inh}}(t)=\sum_{k=-\infty}^{\infty}\varphi_{k}\tilde{d}_{2}(k\nu)\,e^{-ik\nu t}\,.
\end{equation}
Comparing with \eqref{E:dsjfge}, we find that $\mathsf{Q}_{\text{inh}}(t)$ is also a periodic function having the same period of $\varphi_{\text{h}}(\bm{0},t)$ at late times, if we identify $\varphi_{k}\tilde{d}_{2}(k\nu)=g_{k}$ and have waited for a sufficiently long time to let the contributions from the poles in $\tilde{d}_{2}$ drop off. This generalizes the result of the sinusoidal drive case.

On the other hand, if the frequency spectrum of $\varphi_{\text{h}}(\bm{0},t)$ is not like a comb, then $\varphi_{\text{h}}$ is not a periodic function of $t$. We further assume that the spectrum is broad and is a sufficiently continuous function of $\omega$. Practically speaking, in a typical preparation of the coherent state, since only a finite amount of energy and finite spacetime extension are involved, it is much harder to excite the extreme low- and high-frequency modes. Thus, the occupation number, proportional to $\lvert\alpha_{\bm{k}}\rvert^{2}$, is expected to fall off there, and $\alpha_{\omega}$ is expected to decrease rapidly in the low- and the high-frequency ends of the frequency spectrum. Then $\varphi_{\text{h}}(\bm{0},t)$ should be sufficiently well defined by the integral expression \eqref{E:kgeusds}. Hence we expect that $\mathsf{Q}_{\text{inh}}(t)$ is also well behaved. Under this assumption, we observe from \eqref{E:dkjbsss} and \eqref{E:yqeus} that in the limit $t\to\infty$, the asymptotic analysis implies that $\varphi_{\text{h}}(\bm{0},t)$ at least falls off algebraically with $t$.

\begin{figure} 	
	\centering     
	\scalebox{0.5}{\includegraphics{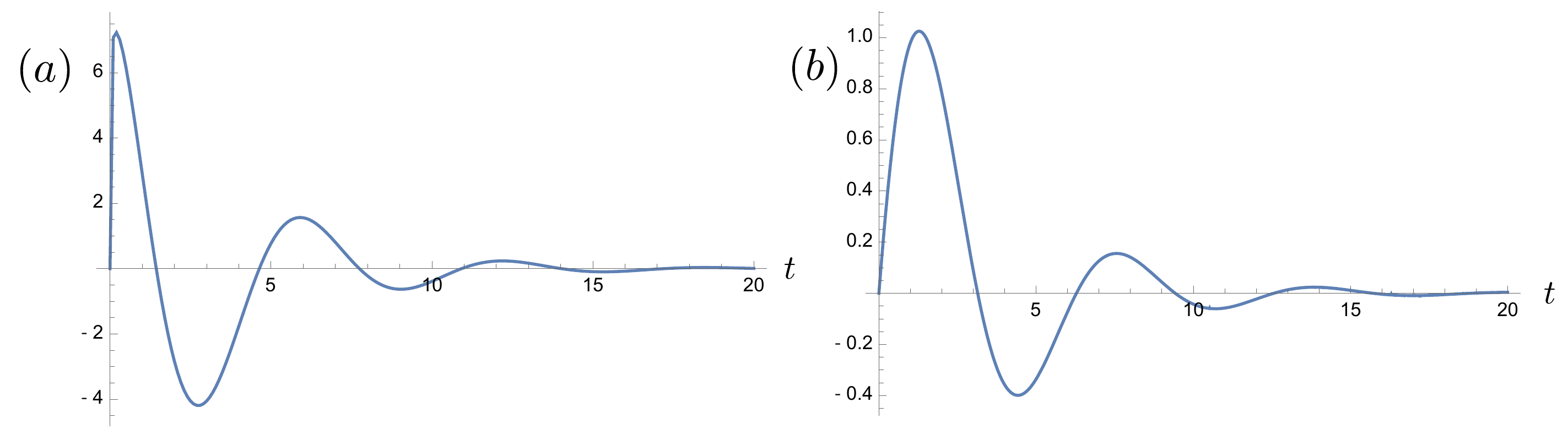}}     
	\caption{The temporal behavior of $\mathsf{Q}_{\text{inh}}(t)$ for two choices of $\alpha_{\omega}$. We have chosen the resonance frequency $\omega_{\textsc{r}}=1$, the damping constant $\gamma=0.2$ and $\epsilon=0.001$. In (a), the coherent parameter is given by \eqref{E:utdfgk1}, and in (b), the coherent parameter has the form in \eqref{E:utdfgk2}}\label{Fi:Qinh} 
\end{figure}

For example, suppose $\alpha_{\omega}$ takes the form (See Fig.~\ref{Fi:Qinh})
\begin{equation}\label{E:utdfgk1}
	\alpha_{\omega}=\frac{\alpha}{\sqrt{\omega}}\,,
\end{equation}
and then \eqref{E:dkjbsss} becomes
\begin{align}\label{E:eklkdgfg}
	\mathsf{Q}_{\text{inh}}(t)&=\frac{8\sqrt{\pi}\gamma}{m}\,e^{-\gamma t}\biggl\{\alpha\Bigl[-i\,\frac{\sin\Omega t}{\Omega\epsilon}-\frac{\Omega\,\cos\Omega t-\gamma\,\sin\Omega t}{\Omega}\,\ln\epsilon+\cdots\Bigr]+\text{C.C.}\biggr\}\,,
\end{align}
where $\cdots$ denotes the cutoff-independent terms, and $\epsilon^{-1}$ is the frequency cutoff, representing the highest energy scale in the coherent state preparation. Since
\begin{equation}\label{E:gksbfh}
	\lim_{t\to\infty}\mathsf{Q}_{\text{inh}}(t)=0\,, 
\end{equation}
the cutoff virtually has no effect at late times. From \eqref{E:kbgeugfd}, the corresponding classical field $\varphi_{\text{h}}(\bm{x},t)$ is given by
\begin{align}
	\varphi_{\text{h}}(\bm{x},t)=\frac{1}{\sqrt{\pi}}\biggl[\frac{\alpha}{r^{2}-(t-i\epsilon)^{2}}+\frac{\alpha^{*}}{r^{2}-(t+i\epsilon)^{2}}\biggr]\,.
\end{align}
It can be verified to satisfy the wave equation once we put $r=\sqrt{x^{2}+y^{2}+z^{2}}$. Another example is 
\begin{equation}\label{E:utdfgk2}
	\alpha_{\omega}=\frac{\alpha}{\sqrt{\omega^{3}}}\,.
\end{equation}
We then have
\begin{align}\label{E:yruebkd}
	\mathsf{Q}_{\text{inh}}(t)&=\frac{8\sqrt{\pi}\gamma}{m}\,\biggl\{\alpha\Bigl[-i\,\frac{1}{(\Omega^{2}+\gamma^{2})t}+e^{-\gamma t}\Bigl(\cdots\Bigr)\Bigr]+\text{C.C.}\biggr\}\,.
\end{align}
Here $\cdots$ represents contributions that are exponentially small at late times as seen in Fig.~\ref{Fi:Qinh}, and again we reach the conclusion as in \eqref{E:gksbfh}. In this case, the classical field $\varphi_{\text{h}}(\bm{x},t)$ has the form
\begin{align}
	\varphi_{\text{h}}(\bm{x},t)&=\frac{\alpha}{\sqrt{\pi}r}\biggl\{\theta(r>\lvert t\rvert)\Bigl[\frac{\pi}{2}-i\,\tanh^{-1}\frac{t}{r}\Bigr]+\theta(\lvert t\rvert-r)\,\Bigl[-i\,\tanh^{-1}\frac{r}{t}\Bigr]\biggr\}+\text{C.C.}\,.
\end{align}
It falls off in the large $t$ limit. 

Finally, we choose a more mundane example for $\varphi_{\text{h}}(\bm{0},t)$ given by a Lorentzian function
\begin{align}
	\varphi_{\text{h}}(\bm{0},t)&=\frac{A}{1+(t/\tau)^{2}}\,.
\end{align}
It yields
\begin{align}	
	\mathsf{Q}_{\text{inh}}(t)&=\frac{A\tau^{2}}{\omega_{\textsc{r}}^{2}t^{2}}+\cdots\,,\qquad\qquad\text{for $t\gg\gamma^{-1}$}\,,
\end{align}
which vanishes at late times. It may be worthwhile to take a look at the critical case that $\varphi_{\text{h}}(\bm{0},t)$ is a constant in time, we obtain
\begin{align}
	\varphi_{\text{h}}(\bm{0},t)&=A\,,&&\Rightarrow&\mathsf{Q}_{\text{inh}}(t)&=\frac{A}{\omega^{2}}\Bigl(1-e^{-\gamma t}\cos\Omega t-\frac{\gamma}{\Omega}\,e^{-\gamma t}\sin\Omega t\Bigr)\to\frac{A}{\omega^{2}}\,.
\end{align}
Note that the frequency spectrum $\varphi_{\text{h}}(\bm{0},t)=A$ is already proportional to $\delta(\omega)/\omega^{\frac{3}{2}}$. Thus we can classify {this} $\varphi_{\text{h}}(\bm{0},t)$ as a periodic function.

\subsection{Three types of late time behavior }

From the previous examples, we may categorize $\varphi_{\text{h}}(\bm{0},t)$ into the following three classes and their combinations:  

A) $\varphi_{\text{h}}(\bm{0},t)$ is a periodic function, so $\mathsf{Q}_{\text{inh}}(t)$ remains nonvanishing as $t\to\infty$. This implies that $\prescript{\alpha}{}{G}_{0,\textsc{H}}^{(\phi)}(\bm{0},s;\bm{0},s')$ continues to be nonstationary according to \eqref{E:rjgfmjnfg}. Thus in this case, the equilibrium state of the reduced internal dynamics of the atom is not likely to exist. 

B) $\varphi_{\text{h}}(\bm{0},t)$ is well behaved by requiring its frequency spectrum to be sufficiently smooth. Then $\varphi_{\text{h}}(\bm{0},t)$ and $\mathsf{Q}_{\text{inh}}(t)$ tend to zero as $t\to\infty$. In this case $\prescript{\alpha}{}{G}_{0,\textsc{H}}^{(\phi)}(\bm{0},s;\bm{0},s')$ reduces to $\prescript{\textsc{vac}}{}{G}_{0,\textsc{H}}^{(\phi)}(\bm{0},s;\bm{0},s')$ at late times, and thus it becomes stationary. We can have an equilibrium state for the reduced $\hat{Q}$ dynamics in this case. 

C) $\varphi_{\text{h}}(\bm{0},t)$ grows unbounded at late times. We do not consider this case because it may not be physically realizable.


\section{energy exchange between the idf of the atom and the field}\label{S:bgvruts} 

We have seen from \eqref{E:lmdius} that two force operators $e\,\hat{\phi}_{\text{h}}(\bm{0},t)$ and $-2m\gamma\,\dot{\hat{Q}}(t)$ account for the energy exchange between the field and the atom. The field's quantum noise force associated with the ubiquitous vacuum fluctuations  drives the atom's internal degree of freedom into  random motion, while the damping force in the atom's internal degree of freedom  disperses the energy  back to the field, and at late times, going in tandem with it. This correlation can be appreciated once we note the damping force depends on the state of the motion via $\dot{\hat{Q}}(t)$, which in turns is governed by the noise force at late times. However, at this stage, it is not clear whether the existence of such correlations is sufficient reason for the equilibration of the reduced dynamics of $\hat{Q}$. This is what we now shall focus on, in examining the role of correlations in the energy exchange between the atom and the field.

The power pumped to the atom by the noise force is defined and given by
\begin{align}
	P_{\xi}(t)&=\frac{e}{2}\langle\bigl\{\hat{\phi}_{\text{h}}(\bm{0},t),\dot{\hat{Q}}(t)\bigr\}\rangle=e\,\varphi_{\text{h}}(\bm{0},t)\dot{\mathsf{Q}}(t)+e^{2}\int_{0}^{t}\!ds\;\frac{\partial}{\partial t}G_{\textsc{R}}^{(Q)}(t-s)\,\prescript{\alpha}{}{G}_{0,\textsc{H}}^{(\phi)}(\bm{0},t;\bm{0},s)\,,
\end{align}
since $\dot{\hat{Q}}_{\text{h}}(t)$ and $\hat{\phi}_{\text{h}}(\bm{0},t)$ are not correlated. With the help of \eqref{E:gldjgfkd}, the power of the noise force becomes
\begin{align}
	P_{\xi}(t)&=e\,\varphi_{\text{h}}(\bm{0},t)\dot{\mathsf{Q}}(t)+e^{2}\int_{0}^{t}\!ds\;\frac{\partial}{\partial t}G_{\textsc{R}}^{(Q)}(t-s)\,\Bigl[\varphi_{\text{h}}(\bm{0},t)\varphi_{\text{s}}(\bm{0},s)+\prescript{\textsc{vac}}{}{G}_{0,\textsc{H}}^{(\phi)}(\bm{0},t;\bm{0},s)\Bigr]\notag\\
	&=\mathcal{P}_{\xi}(t)+P^{0}_{\xi}(t)\,,
\end{align}
where
\begin{align}
	\mathcal{P}_{\xi}(t)&=e\,\varphi_{\text{h}}(\bm{0},t)\dot{\mathsf{Q}}(t)+e^{2}\int_{0}^{t}\!ds\;\frac{\partial}{\partial t}G_{\textsc{R}}^{(Q)}(t-s)\,\varphi_{\text{h}}(\bm{0},t)\varphi_{\text{s}}(\bm{0},s)\,,\\
	P^{0}_{\xi}(t)&=e^{2}\int_{0}^{t}\!ds\;\frac{\partial}{\partial t}G_{\textsc{R}}^{(Q)}(t-s)\,\prescript{\textsc{vac}}{}{G}_{0,\textsc{H}}^{(\phi)}(\bm{0},t;\bm{0},s)\,.\label{E:dgkfbdfg}
\end{align}
The power $\mathcal{P}_{\xi}(t)$ gives the contribution due to the classical, deterministic component of $P_{\xi}(t)$. That is,  the power delivered by the mean force $e\,\varphi_{\text{h}}(\bm{0},t)$ in the classical equation of motion \eqref{E:lmdius2}
\begin{equation*}
	\ddot{\mathsf{Q}}(t)+2\gamma\,\dot{\mathsf{Q}}(t)+\omega_{\textsc{r}}^{2}\mathsf{Q}(t)=\frac{e}{m}\,\varphi_{\text{h}}(\bm{z},t)\,.
\end{equation*}
It is straightforward to see that the power delivered by $e\,\varphi_{\text{h}}(\bm{0},t)$ is
\begin{align}
	e\,\varphi_{\text{h}}(\bm{0},t)\dot{\mathsf{Q}}(t)=e\,\varphi_{\text{h}}(\bm{0},t)\Bigl[\dot{\mathsf{Q}}_{\text{h}}(t)+e\int_{0}^{t}\!ds\;\frac{\partial}{\partial t}G_{\textsc{R}}^{(Q)}(t-s)\,\varphi_{\text{h}}(\bm{0},s)\Bigr]
\end{align}
by \eqref{E:iefjdn}, and is exactly $\mathcal{P}_{\xi}(t)$. This classical power can be nonzero even at late time, when the classical force $e\,\varphi_{\text{h}}(\bm{0},t)$ is periodic, but it can vanish, not necessarily exponentially, if the classical force has a sufficiently smooth spectrum, according to our earlier discussions.

Eq.~\eqref{E:dgkfbdfg}, in contrast, has a quantum-mechanical origin. The driving force is caused by the zero-point fluctuations of the field, and the component of the velocity of the atom's internal dynamics involved in $P^{0}_{\xi}(t)$ is also governed by the same quantum field fluctuations. This is   most transparently seen by the equation of motion \eqref{E:lmdius3} for $\hat{q}$ 
\begin{equation*}
	\ddot{\hat{q}}(t)+2\gamma\,\dot{\hat{q}}(t)+\omega_{\textsc{r}}^{2}\hat{q}(t)=\frac{e}{m}\,\Bigl[\hat{\phi}_{\text{h}}(\bm{0},t)-\varphi_{\text{h}}(\bm{0},t)\Bigr]\,.
\end{equation*}
We see the power input by $e[\hat{\phi}_{\text{h}}(\bm{0},t)-\varphi_{\text{h}}(\bm{0},t)]$ is exactly $P^{0}_{\xi}(t)$. This quantum power, as shown in~\cite{QRad}, will always reach a time-independent constant for $t\gg\gamma^{-1}$.

The damping force $-2m\gamma\,\dot{\hat{Q}}(t)$ will dissipate the energy of the atom's internal dynamics at a rate given by
\begin{align}
	P_{\gamma}(t)&=-2m\gamma\,\langle\dot{\hat{Q}}^{2}(t)\rangle=\mathcal{P}_{\gamma}(t)+P^{0}_{\gamma}(t)\,,
\end{align}
where likewise we define
\begin{align}
	\mathcal{P}_{\gamma}(t)&=-2m\gamma\,\biggl[\dot{\mathsf{Q}}_{\text{h}}(t)+\,e\,\dot{\mathsf{Q}}_{\text{h}}(t)\int_{0}^{t}\!ds\;\frac{\partial}{\partial t}G_{\textsc{R}}^{(Q)}(t-s)\,\varphi_{\text{h}}(\bm{0},s)\biggr]^{2}\,,\\
	P^{0}_{\gamma}(t)&=-2m\gamma\,\langle\dot{\hat{q}}_{\text{h}}^{2}(t)\rangle-2m\gamma\,e^{2}\int_{0}^{t}\!dsds'\;\frac{\partial}{\partial t}G_{\textsc{R}}^{(Q)}(t-s)\,\frac{\partial}{\partial t}G_{\textsc{R}}^{(Q)}(t-s')\,\prescript{\textsc{vac}}{}{G}_{0,\textsc{H}}^{(\phi)}(\bm{0},s;\bm{0},s')\,.\label{E:dff}
\end{align}
This dissipation power consists of two contributions, one of a purely classical origin and one of a purely quantum origin. We will come to this point when we discuss the radiation. For the interacting Gaussian systems under consideration, the classical and the quantum dynamics are decoupled, so we have a clear distinction in the components of the energy flow between the two subsystems. Here the mean dynamics exclusively follows a classical description, but this is not always true~\cite{Ja74}.

It has been shown~\cite{QTD1,QRad,NLFDR,CPR2D,CPR4D} that the quantum power $P^{0}_{\gamma}(t)$ also approaches a constant at late times and is then balanced by $P^{0}_{\xi}(t)$
\begin{equation}\label{E:ejtkbgf}
	\lim_{t\gg\gamma^{-1}}P^{0}_{\xi}(t)+P^{0}_{\gamma}(t)=0\,.
\end{equation}
In particular
\begin{align}
	P^{0}_{\xi}(\infty)&=e^{2}\int^{\infty}_{-\infty}\!\frac{d\omega}{2\pi}\;\operatorname{sgn}(\omega)\,\frac{\omega^{2}}{4\pi}\,\operatorname{Im}\tilde{G}_{\textsc{R}}^{(Q)}(\omega)\,,&P^{0}_{\gamma}(\infty)&=-P^{0}_{\xi}(\infty)\,.\label{E:dgnksdger}
\end{align}
Eq.~\eqref{E:ejtkbgf} does not depend  on the values of $\varphi_{\text{h}}(t)$ and $\mathsf{Q}(t)$ because from previous discussions we have noticed that the dynamics of $\mathsf{Q}$ and $\hat{q}$ are independent. In the case $\varphi_{\text{h}}(t)\to0$ when $t\to\infty$, the reduced dynamics of $\hat{Q}$ can equilibrate at late times, as if the internal dynamics of the atom is coupled to the vacuum fluctuations of the scalar field $\hat{\phi}_{\text{h}}$. Nonetheless if $\varphi_{\text{h}}(t)$ is periodic, then in general
\begin{equation}
	\lim_{t\gg\gamma^{-1}}\mathcal{P}_{\xi}(t)+\mathcal{P}_{\gamma}(t)\neq0
\end{equation}
as shown in an earlier example, so equilibration does not exist in this case.  In addition, we recall  that the balance between energy exchange alone does not guarantee the existence of  an equilibrium state. This can be seen in the case of a classical damped oscillator driven by a single-mode sinusoidal force, whose driving frequency is equal to the physical frequency of the oscillator. We also need stationarity of the correlation function of $\hat{Q}$ or the constancy of the powers associated with the energy exchange at late times.

A short summary of what we have learned from the reduced dynamics of the internal degree of freedom of the atom is perhaps useful here.  When an atom is coupled to a quantum field initially in a coherent state,  their interaction will entice the atom to emit   radiation into the  field, whence  the quantum field  has two components: the original free field and the  radiation field generated by the atom. The free-field component plays the role of a driving force acting on the internal dynamics of the atom, while the  reaction from the radiation field gives rise to a damping term in  the atom's internal motion. When the free field is initially in the coherent state, it possesses a non-zero mean field. This  introduces a nonvanishing mean dynamics to the atom's internal degree of freedom, and often dominates over the contributions from the quantum fluctuations.  Quantum fluctuations in the atom's internal dynamics  contain both the intrinsic fluctuations of the idf and the induced one from the quantum field. 

The late-time behavior of the free mean field leads to different outcomes.  We are interested in only two types of free mean field dynamics: either periodic in time or decreasing with time to zero. When the mean field is periodic in time, the internal dynamics of the atom will  also be periodic at late times. Thus if the coherent parameter of the free field is much greater than one, the classical mean dynamics of the atom's internal motion always dominates over the fluctuation dynamics, which is governed by the vacuum fluctuations of the field. In this case, the correlation function of the atom's internal dynamics is never stationary in time and  will not reach equilibrium. The energy pumped into the atom by the free mean field per unit time is not in sync with the rate of the energy dissipated by the atom's internal degree of freedom, so there is net energy exchange between the atom and the field throughout the evolution.

In contrast, if the mean field vanishes at late times, the atom's mean internal dynamics  likewise approaches zero at late times. Thus during the transient moment, even though the internal motion is still dominated by the classical mean dynamics, it will in the end be controlled by the quantum fluctuation dynamics. This implies that the correlation function of the atom's internal dynamics gradually turns stationary in time because its classical nonstationary component vanishes by then. The internal dynamics will equilibrate such that the energy exchange between the atom and the field comes to a balance.

In the current setting, since the mean dynamics of the internal degree of freedom of the atom is decoupled from its fluctuation dynamics, we can examine their behaviors separately.  We have shown that the fluctuation dynamics always comes to equilibration, independent of the mean dynamics, so the late-time behavior of the atom's reduced dynamics is determined by its mean component. If the mean dynamics does not vanish at late time, then the atom's internal motion will not equilibrate. On the other hand, if the mean dynamics diminishes asymptotically, then the fluctuation dynamics will ensure the atom's reduced dynamics will settle down to an equilibrium state. This analysis showcases the differences between the classical deterministic dynamics and the its quantum fluctuating dynamics of the atom's internal degree of freedom.

\section{radiation flux at spatial infinity}\label{S:eobgd}


In the previous section, we have shown that the quantum field pumps energy into the atom's internal dynamics, and field fluctuations imparts a random component in the motion of its internal degrees of freedom.  The atom in turn will radiate energy back to its surrounding field in the form of a radiation field. When the equilibrium condition is established between the atom's idf and the field, two questions of interest may arise: 1) If the field is initially in the vacuum state  where does the pumping energy come from?  2) Where does the radiated energy go?  Superficially it may sound like a perpetual motion --  extracting the vacuum energy for practical use. In fact these two questions are both sides of the same coin.    We shall begin with this puzzle and  address this issue  from the viewpoint of radiation emitted by the atom.

The radiation power of the scalar field at an observation point $x$ far away from the atom located at $z$ is given by 
\begin{align}\label{E:fhjgdfjsd}
	\frac{dW_{\textsc{rad}}}{d\tau}=-\int\!d\Omega\;r^{2}n^{\mu}\langle\, \wr\, T_{\mu\nu}(x)\,\wr\,\rangle v^{\nu}(\tau_{-})\,,
\end{align}
where $v^{\mu}=dz^{\mu}/d\tau$ is the external four-velocity of the harmonic atom and $\tau_{-}$ is the retarded time.  Since we are interested in a stationary atom in flat spacetime, the proper time of the atom $\tau$ will be synonymous with the coordinate time $t$. The spacelike unit normal vector $n^{\mu}$ specifies the outward radial direction, and the distance $r$ between the observation point and the atom is given by the projection
\begin{equation}
	r=n_{\mu}\bigl[x^{\mu}-z^{\mu}(\tau_{-})\bigr]\,.
\end{equation}
This defines a spherical shell centered at the atom at the retarded time $\tau_{-}$. In \eqref{E:fhjgdfjsd}, we introduce a shorthand notation\footnote{We take this opportunity to point out a misnomer in~\cite{QRad}: the `normal-ordered' stress tensor operator should be the stress tensor operator \eqref{E:kfgjbdrst} defined here. The results in~\cite{QRad} remain valid despite the incorrect notation.} $\,\wr\,\hat{T}_{\mu\nu}(x^{\alpha})\,\wr\,$, defined by 
\begin{equation}\label{E:kfgjbdrst}
	\wr\,\hat{T}_{\mu\nu}(x)\,\wr=\hat{T}_{\mu\nu}(x)-\hat{T}^{(h)}_{\mu\nu}(x)\,.
\end{equation}
That is, we subtract from the full energy-momentum stress tensor operator $\hat{T}_{\mu\nu}(x)$ the component $\hat{T}_{\mu\nu}^{(h)}(x)$, which is the corresponding stress tensor operator for the free in-field $\hat{\phi}_{\text{h}}(x)$ and does not contain any field generated by the atom. The stress tensor of a classical linear massless scalar field is given by
\begin{equation}
	T_{\mu\nu}=-\frac{2}{\sqrt{-\eta}}\frac{\delta S_{\textsc{f}}}{\delta \eta^{\mu\nu}}=\partial_{\mu}\phi\,\partial_{\nu}\phi-\frac{1}{2}g_{\mu\nu}\,g^{\alpha\beta}\partial_{\alpha}\phi\,\partial_{\beta}\phi\,.
\end{equation}
When the field is promoted to an operator we may conveniently express the expectation value of $\,\wr\,\hat{T}_{\mu\nu}(x^{\alpha})\,\wr\,$ by the Hadamard function of the scalar field in the coincident limit
\begin{equation}
	\langle \,\wr\, T_{\mu\nu}(x)\,\wr\,\rangle=\lim_{x'\to x}\biggl\{\frac{\partial^{2}}{\partial x^{\mu}\partial x'^{\nu}}-\frac{1}{2}\,g^{\mu\nu}g^{\alpha\beta}\frac{\partial^{2}}{\partial x^{\alpha}\partial x'^{\beta}}\biggr\}\Bigl[G_{\textsc{H}}^{(\phi)}(x,x')-G_{0,\textsc{H}}^{(\phi)}(x,x')\Bigr]\,,\label{E:siushgfd}
\end{equation}
so that
\begin{equation}\label{E:eotuhgfd}
	\frac{dW_{\textsc{rad}}}{d\tau}=-\lim_{x'\to x}\int\!d\Omega_{\bm{k}}\;r^{2}\,\frac{\partial^{2}}{\partial r\partial t'}\Bigl[G_{\textsc{H}}^{(\phi)}(x,x')-G_{0,\textsc{H}}^{(\phi)}(x,x')\Bigr]\,,
\end{equation}
where $d\Omega_{\bm{k}}$ is the solid angle subtended over a spherical shell centered at the atom at the retarded time $\tau_{-}$. We are interested in this radiation power at a distance sufficiently far away from the atom. By `sufficiently  far', we refer to a distance where the dominant contribution can be identified and is independent of this distance.

The Hadamard function, the expectation value of the anti-commutator, of the scalar field $G_{\textsc{H}}^{(\phi)}(x,x')$ is defined by
\begin{equation}
	G_{\textsc{H}}^{(\phi)}(x,x')=\frac{1}{2}\langle\bigl\{\hat{\phi}(x),\hat{\phi}(x')\bigr\}\rangle\,,
\end{equation}
where in the Heisenberg picture the expectation value is computed with respect to the initial state of the total system, and $G_{0,\textsc{H}}^{(\phi)}(x,x')$ is the corresponding Hadamard function for the free field $\hat{\phi}_{\text{h}}$. Here we would like to emphasize again the difference between $G_{\textsc{H}}^{(\phi)}(x,x')$ and $G_{0,\textsc{H}}^{(\phi)}(x,x')$; the latter is the Hadamard function of the free field while the former is that of the full field, the sum of the free field and the field radiated by the atom. By radiation field, we mean the component of the field generated by the motion of the atom's internal degree of freedom. Finally, in contrast to the retarded Green's function of the scalar field which is state-independent, the behavior of the Hadamard function depends on the choice of the initial state of the field. Thus \eqref{E:fhjgdfjsd} gives the total radiated power of the field emitted by the atom at the retarded time $\tau_{-}$.

Since the Hadamard function $G_{\textsc{H}}^{(\phi)}(x,x')$ of the scalar field plays a central role, hereafter we will delineate its structure. From \eqref{E:lgndlg3}, we find that the scalar in general will take on the form
\begin{align}\label{E:urueuya}
	\hat{\phi}(x)=\hat{\phi}_{\text{h}}(x)+\hat{\phi}_{\text{inh}}(x)=\hat{\phi}_{\text{h}}(x)+\hat{\phi}_{\textsc{tr}}(x)+\hat{\phi}_{\textsc{r}}(x)\,,
\end{align}
where
\begin{equation}
	\hat{\phi}_{\text{inh}}(x)=e\int_{0}^{t}\!ds\;G_{0,\textsc{R}}^{(\phi)}(\bm{x},t;\bm{z},s)\,\hat{Q}(s)\,.\label{E:baosidfh} 
\end{equation}
The homogeneous solution  $\hat{\phi}_{\text{h}}$  corresponds to the free field, not affected by the internal degree of freedom $\hat{Q}$ of the atom. The radiated field $\hat{\phi}_{\text{inh}}(x)$ emitted by the atom contains two components denoted by
\begin{align}
	\hat{\phi}_{\textsc{tr}}(x)&=e\int\!dt'\;G_{0,\textsc{R}}^{(\phi)}(\bm{x},t;\bm{z},t')\,\hat{Q}_{\text{h}}(t')\,,\\
	\hat{\phi}_{\textsc{r}}(x)&=e^{2}\int\!dt'\;G_{0,\textsc{R}}^{(\phi)}(\bm{x},t;\bm{z},t')\int\!dt''\;G_{\textsc{R}}^{(Q)}(t'-t'')\hat{\phi}_{\text{h}}(\bm{z},t'')\,,
\end{align}
where we have decomposed $\hat{Q}$ according to \eqref{E:gkdfgjsd}. The operator $\hat{Q}_{\text{h}}$ represents the transient motion of the internal degree of freedom of the atom. Since this transient dynamics decays to zero exponentially fast with respect to the relaxation time $\gamma^{-1}$, the field operator $\hat{\phi}_{\textsc{tr}}$ associated with this component is usually ignored at late times, even in the vicinity of the atom. The field $\hat{\phi}_{\textsc{r}}$ on the other hand is produced due to the resonant motion of $\hat{Q}$ driven by the free scalar field $\hat{\phi}_{\text{h}}$. Since the resonant motion of $\hat{Q}$, the particular solution of \eqref{E:bgweirusd}, will survive at late times, the field $\hat{\phi}_{\textsc{r}}$ will be emitted unbated by the atom. More importantly, this implies that at late times the dynamics of $\hat{\phi}_{\textsc{r}}$ will be correlated with the dynamics of $\hat{\phi}_{\text{h}}$ and the initial state of the field.

The Hadamard function formed by \eqref{E:urueuya} can be classified into three contributions
\begin{align}
	G_{\textsc{H}}^{(\phi)}(x,x')&=G_{0,\textsc{H}}^{(\phi)}(x,x')+\Bigl[\frac{1}{2}\langle\bigl\{\hat{\phi}_{\text{h}}(x),\hat{\phi}_{\textsc{r}}(x')\bigr\}\rangle+\frac{1}{2}\langle\bigl\{\hat{\phi}_{\textsc{r}}(x),\hat{\phi}_{\text{h}}(x')\bigr\}\rangle\Bigr]+\frac{1}{2}\langle\bigl\{\hat{\phi}_{\textsc{r}}(x),\hat{\phi}_{\textsc{r}}(x')\bigr\}\rangle\,,\label{E:gksgfer}
\end{align}
at late times. There are no cross terms like $\langle\bigl\{\hat{\phi}_{\text{h}}(x),\hat{\phi}_{\textsc{tr}}(x')\bigr\}\rangle$, $\langle\bigl\{\hat{\phi}_{\textsc{tr}}(x),\hat{\phi}_{\textsc{r}}(x')\bigr\}\rangle$ and $\langle\bigl\{\hat{\phi}_{\textsc{tr}}(x),\hat{\phi}_{\textsc{tr}}(x')\bigr\}\rangle$ because they are exponentially suppressed at late times. Thus we have
\begin{align}
	\frac{1}{2}\langle\bigl\{\hat{\phi}_{\text{h}}(x),\hat{\phi}_{\textsc{r}}(x')\bigr\}\rangle&=e^{2}\int^{t'}_{0}\!ds\;G_{0,\textsc{R}}^{(\phi)}(\bm{x}',t';\bm{z},s)\int_{0}^{s}\!ds'\;G_{\textsc{R}}^{(Q)}(s-s')\,G_{0,\textsc{H}}^{(\phi)}(\bm{x},t;\bm{z},s')\,,\label{E:keoruhs1}\\
	\frac{1}{2}\langle\bigl\{\hat{\phi}_{\textsc{r}}(x),\hat{\phi}_{\textsc{r}}(x')\bigr\}\rangle&=e^{4}\int_{0}^{t}\!ds_{1}\!\int_{0}^{t'}\!ds_{1}'\;G_{0,\textsc{R}}^{(\phi)}(\bm{x},t;\bm{z},s_{1})\,G_{0,\textsc{R}}^{(\phi)}(\bm{x}',t';\bm{z},s'_{1})\label{E:keoruhs3}\\
	&\qquad\qquad\qquad\times\int_{0}^{s_{1}}\!ds_{2}\!\int_{0}^{s'_{1}}\!ds'_{2}\;G_{\textsc{R}}^{(Q)}(s_{1}-s_{2})\,G_{\textsc{R}}^{(Q)}(s'_{1}-s'_{2})\,G_{0,\textsc{H}}^{(\phi)}(\bm{z},s_{2};\bm{z},s'_{2})\,.\notag
\end{align}
Here $G_{\textsc{R}}^{(Q)}$ is the retarded Green's function of the full interacting oscillator, satisfying \eqref{E:dgksdhser}.

Since we prepare the quantum field initially in a coherent state, the Hadamard functions $\prescript{\alpha}{}{G}_{0,\textsc{H}}^{(\phi)}(\bm{z},s;\bm{z}',s')$ of the free quantum scalar field $\hat{\phi}_{\text{h}}$ in \eqref{E:keoruhs1}--\eqref{E:keoruhs3} can be separated into the classical and the quantum contributions
\begin{align}
	\prescript{\alpha}{}{G}_{0,\textsc{H}}^{(\phi)}(\bm{z},s;\bm{z}',s')=\varphi_{\text{h}}(\bm{z},s)\varphi_{\text{h}}(\bm{z}',s')+\prescript{\textsc{vac}}{}{G}_{0,\textsc{H}}^{(\phi)}(\bm{z},s;\bm{z}',s')\,,\label{E:gbfhfgfd}
\end{align}
We then can write the righthand sides of \eqref{E:keoruhs1}--\eqref{E:keoruhs3} into similar decompositions,
\begin{align}
	\frac{1}{2}\langle\bigl\{\hat{\phi}_{\text{h}}(x),\hat{\phi}_{\textsc{r}}(x')\bigr\}\rangle&=\varphi_{\text{h}}(x)\varphi_{\textsc{r}}(x')+\frac{1}{2}\langle\bigl\{\hat{\phi}_{\text{h}}(x),\hat{\phi}_{\textsc{r}}(x')\bigr\}\rangle_{0}\,,\label{E:keoruhs4}\\
	\frac{1}{2}\langle\bigl\{\hat{\phi}_{\textsc{r}}(x),\hat{\phi}_{\textsc{r}}(x')\bigr\}\rangle&=\varphi_{\textsc{r}}(x)\varphi_{\textsc{r}}(x')+\frac{1}{2}\langle\bigl\{\hat{\phi}_{\textsc{r}}(x),\hat{\phi}_{\textsc{r}}(x')\bigr\}\rangle_{0}\,,\label{E:keoruhs6}
\end{align}
where the subscript $0$ denotes the field state used for the expectation value is the vacuum state, and 
\begin{equation}
	\varphi_{\textsc{r}}(x)=e^{2}\int_{0}^{t}\!ds\;G_{0,\textsc{R}}^{(\phi)}(\bm{x},t;\bm{z},s)\int_{0}^{s}\!ds'\;G_{\textsc{R}}^{(Q)}(s-s')\,\varphi_{\text{h}}(\bm{z},s)=e\int_{0}^{t}\!ds\;G_{0,\textsc{R}}^{(\phi)}(\bm{x},t;\bm{z},s)\,\mathsf{Q}_{\text{inh}}(s)\,.
\end{equation}
Here we recall that $G_{0,\textsc{R}}^{(\phi)}(\bm{z},s;\bm{z}',s')$ is state-independent, so there is no difference between $\prescript{\alpha}{}{G}_{0,\textsc{R}}^{(\phi)}(\bm{z},s;\bm{z}',s')$ and $\prescript{\textsc{vac}}{}{G}_{0,\textsc{R}}^{(\phi)}(\bm{z},s;\bm{z}',s')$.

Alternative to writing $\hat{\phi}_{\text{inh}}(x)$ as the sum of $\hat{\phi}_{\textsc{tr}}(x)$ and $\hat{\phi}_{\textsc{r}}(x)$, we use $\hat{q}=\hat{Q}-\mathsf{Q}$ to break $\hat{\phi}_{\text{inh}}(x)$ in \eqref{E:baosidfh} into contributions of the classical and the quantum natures,
\begin{equation}\label{E:lfkgjerts}
	\hat{\phi}_{\text{inh}}(x)=\varphi_{\text{inh}}(x)+\bigl[\hat{\phi}_{\text{inh}}(x)-\varphi_{\text{inh}}(x)\bigr]\,,
\end{equation}
where
\begin{align}
	\varphi_{\text{inh}}(x)&=e\int_{0}^{t}\!ds\;G_{0,\textsc{R}}^{(\phi)}(\bm{x},t;\bm{z},s)\,\mathsf{Q}(s)\label{E:kdfgbvkfg}\\
			&=e\int_{0}^{t}\!ds\;G_{0,\textsc{R}}^{(\phi)}(\bm{x},t;\bm{z},s)\,\mathsf{Q}_{\text{h}}(s)+e^{2}\int_{0}^{t}\!ds\;G_{0,\textsc{R}}^{(\phi)}(\bm{x},t;\bm{z},s)\int_{0}^{s}\!ds'\;G_{\textsc{R}}^{(Q)}(s-s')\,\varphi_{\text{h}}(\bm{z},s')\,,\notag
\end{align}
corresponds to the classical radiation due to the classical motion $\mathsf{Q}$,  the first term being classical transient radiation which becomes negligibly small at late times. The  terms inside the square brackets in \eqref{E:lfkgjerts} result  from the fluctuations $\hat{q}$,
\begin{equation}
	\hat{\phi}_{\text{inh}}(x)-\varphi_{\text{inh}}(x)=e\int_{0}^{t}\!ds\;G_{0,\textsc{R}}^{(\phi)}(\bm{x},t;\bm{z},s)\,\hat{q}(s)\,.
\end{equation}
This is the source of  quantum radiation. It includes both the intrinsic quantum fluctuations $\hat{q}_{\text{h}}$ of $\hat{Q}$, which decays exponentially with time, and the induced quantum fluctuations by the vacuum fluctuations of the free field $\hat{\phi}_{\text{h}}$.

The decompositions in \eqref{E:keoruhs4} and \eqref{E:keoruhs6} imply that the radiation power of the quantum scalar field can be accordingly written into two distinct components
\begin{equation}
	\frac{dW_{\textsc{rad}}}{d\tau}=\frac{d\mathsf{W}_{\textsc{rad}}}{d\tau}+\frac{dw_{\textsc{rad}}}{d\tau}\,,
\end{equation}
where $\mathsf{W}_{\textsc{rad}}$ is ultimately  the contribution caused by the mean field  when the quantum scalar field is initially prepared in the coherent state, while $w_{\textsc{rad}}$ is the contribution from the accompanying vacuum fluctuations of the field. Here we focus on the mean field component $d\mathsf{W}_{\textsc{rad}}/d\tau$ because the other vacuum contribution is already studied in our first paper~\cite{QRad}, but we will include it to show the essential features of both for comparison.

At late times, the classical contributions in \eqref{E:keoruhs4} and \eqref{E:keoruhs6} can be combined into
\begin{align}
	\varphi_{\text{h}}(x)\varphi_{\textsc{r}}(x')+\varphi_{\text{h}}(x')\varphi_{\textsc{r}}(x)+\varphi_{\textsc{r}}(x)\varphi_{\textsc{r}}(x')\,.
\end{align}
at the observation point sufficiently far away from the radiating atom. The corresponding classical stress tensor component is then given by
\begin{align}
	\lim_{x'\to x}\frac{\partial^{2}}{\partial r\partial t'}\Bigl[\varphi_{\text{h}}(x)\varphi_{\textsc{r}}(x')+\varphi_{\text{h}}(x')\varphi_{\textsc{r}}(x)+\varphi_{\textsc{r}}(x)\varphi_{\textsc{r}}(x')\Bigr]\,,\label{E:gkdbfgdf}
\end{align}
as seen from \eqref{E:eotuhgfd}, which can be further reduced to
\begin{equation}
	\Bigl[\frac{\partial}{\partial r}\varphi_{\text{h}}(\bm{x},t)-\frac{\partial}{\partial t}\varphi_{\text{h}}(\bm{x},t)-\frac{\partial}{\partial t}\varphi_{\textsc{r}}(\bm{x},t)\Bigr]\frac{\partial}{\partial t}\varphi_{\textsc{r}}(\bm{x},t)\,.\label{E:kgfhv}
\end{equation}
The first two terms inside the square brackets, corresponding to the contributions of $\varphi_{\text{h}}(x)\varphi_{\textsc{r}}(x')+\varphi_{\text{h}}(x')\varphi_{\textsc{r}}(x)$, are given by
\begin{align}
	\frac{\partial}{\partial r}\varphi_{\text{h}}(\bm{x},t)-\frac{\partial}{\partial t}\varphi_{\text{h}}(\bm{x},t)&=4\sqrt{\pi}\int_{0}^{\infty}\!d\omega\;\omega^{\frac{3}{2}}\Bigl\{\alpha_{\omega}\,\widetilde{G}_{0,\textsc{R}}^{(\phi)}(\mathbf{r};\omega)\,e^{-i\omega t}+\text{C.C.}\Bigr\}\,,\label{E:oeiufhd}
\end{align}
since
\begin{align}
	\varphi_{\text{h}}(\bm{x},t)&=\frac{1}{\sqrt{\pi}r}\int_{0}^{\infty}\!d\omega\;\omega^{\frac{1}{2}}\Bigl[\alpha_{\omega}\,\sin\omega r\,e^{-i\omega t}+\text{C.C.}\Bigr]\,,\\
	\varphi_{\textsc{r}}(\bm{x},t)&=\frac{e^{2}}{\sqrt{\pi}}\int_{0}^{\infty}\!d\omega\;\omega^{\frac{3}{2}}\Bigl[\alpha_{\omega}\,L_{\omega}(\bm{x},t)+\text{C.C.}\Bigr]\,,
\end{align}
at late times. Here we have placed the atom at the origin of the external coordinate system, so $r=\lvert\bm{x}\rvert$, $r'=\lvert\bm{x}'\rvert$ and we have assigned $e^{2}=8\pi\gamma m$, and $L_{\omega}(\bm{x},t)=\theta(t-r)\,\widetilde{G}_{\textsc{R}}^{(Q)}(\omega)\widetilde{G}_{0,\textsc{R}}^{(\phi)}(r;\omega)\,e^{-i\omega t}$. The other factor in \eqref{E:kgfhv} has the form
\begin{align}
	-\frac{\partial}{\partial r}\varphi_{\textsc{r}}(\bm{x},t)=\frac{\partial}{\partial t}\varphi_{\textsc{r}}(\bm{x},t)&=\theta(t-r)\,\frac{e^{2}}{\sqrt{\pi}}\int_{0}^{\infty}\!d\omega\;\omega^{\frac{5}{2}}\Bigl[-i\,\alpha_{\omega}\,\widetilde{G}_{\textsc{R}}^{(Q)}(\omega)\widetilde{G}_{0,\textsc{R}}^{(\phi)}(r;\omega)\,e^{-i\omega t}+\text{C.C.}\Bigr]\,,\label{E:dkgsghfe}
\end{align}
Here we have used 
\begin{align}
	\widetilde{G}_{0,\textsc{R}}^{(\phi)}(\mathbf{r};\omega)&=\frac{e^{+i\omega r}}{4\pi r}\,,&\widetilde{G}_{\textsc{R}}^{(Q)}(\omega)&=\frac{1}{-\omega^{2}-i\,2\gamma\omega+\omega_{\textsc{r}}^{2}}\,,
\end{align}
to re-write the square brackets in \eqref{E:kgfhv} as 
\begin{align}
	&\quad\frac{\partial}{\partial r}\varphi_{\text{h}}(\bm{x},t)-\frac{\partial}{\partial t}\varphi_{\text{h}}(\bm{x},t)-\frac{\partial}{\partial t}\varphi_{\textsc{r}}(\bm{x},t)\notag\\
	&=4m\sqrt{\pi}\int_{0}^{\infty}\!d\omega\;\omega^{\frac{3}{2}}\bigl(\omega^{2}_{\textsc{r}}-\omega^{2}\bigr)\Bigl\{\alpha_{\omega}\,\widetilde{G}_{\textsc{R}}^{(Q)}(\omega)\,\widetilde{G}_{0,\textsc{R}}^{(\phi)}(\mathbf{r};\omega)\,e^{-i\omega t}+\text{C.C.}\Bigr\}\,.\label{E:gnksdjfg}
\end{align}

In order to determine whether the component of the classical stress tensor \eqref{E:gkdbfgdf} constitutes the energy flux far away from the atom at late times, we will examine the temporal behavior of \eqref{E:dkgsghfe} and \eqref{E:gnksdjfg} in the large $t$ limit. According to the discussion in the previous section, if the free mean field $\varphi_{\text{h}}(\bm{0},t)$ is a periodic function of $t$, then comparing \eqref{E:kbgeugfd} with \eqref{E:dsjfge}, we find
\begin{align}
	\frac{1}{\sqrt{\pi}}\,\omega^{\frac{3}{2}}\,\alpha_{\omega}=\frac{\tilde{g}(\omega)}{2\pi}=\frac{\tilde{g}_{\tau}(\omega)}{\tau}\sum_{k=-\infty}^{\infty}\delta(\omega-\frac{2k\pi}{\tau})\,.
\end{align}
This implies that \eqref{E:gnksdjfg} will likewise manifest periodicity at late times after the contributions from the poles of $\widetilde{G}_{\textsc{R}}^{(Q)}$ have dwindled off. Thus in this case the radiation flux passing through a large spherical shell centered at the atom remains nonzero at late times. It is consistent with the results we  found for the internal dynamics of the atom. Since $\varphi_{\textsc{r}}$ in \eqref{E:kdfgbvkfg} can be identified as a classical potential equivalent to the Li\'enard–Wiechert potential in classical electrodynamics, the derivations of the Larmor's formula and the self-force follow suit, which we will not dwell on any further.

On the other hand, if the free mean field $\varphi_{\text{h}}(\bm{0},t)$ falls off to zero at late times $t\gg\gamma^{-1}$, then both $\varphi_{\text{h}}(\bm{x},t)$ and $\varphi_{\textsc{r}}(\bm{x},t)$ will be vanishingly small at even later times $t\gg\gamma^{-1}+r$. Here we stress that the observation point is not literally at spatial infinity from the atom. We only need it to be sufficiently large. Thus, the  classical radiation fluxes will diminish with time, so that in the end it is no longer to be dominant factors. The quantum radiation fluxes, summarized below, still actively leave and enter a large fictitious spherical shell centered at the atom, except that when equilibration is accomplished, the incoming flux is equal to the outgoing flux. Comparing with the former periodic case, we may trace the back-and-forth radiation flux to the periodic behavior of the free mean field.

At this point, it is interesting to recall how the radiation fluxes due to the fluctuation dynamics of the atom's internal degree freedom can possibly come to equilibrium.  This was treated in~\cite{QRad}. We will use the component $\prescript{\textsc{vac}}{}{G}_{0,\textsc{H}}^{(\phi)}(\bm{z},s;\bm{z}',s')$ in \eqref{E:gbfhfgfd} as the contrasting example. Here we shall  summarize the results in~\cite{QRad} and show how they enter in the present case. In this case, we find
\begin{align}
	\frac{1}{2}\langle\bigl\{\hat{\phi}_{\text{h}}(x),\hat{\phi}_{\textsc{r}}(x')\bigr\}\rangle_{0}&=e^{2}\int^{\infty}_{-\infty}\!\frac{d\omega}{2\pi}\;\prescript{\textsc{vac}}{}{\tilde{G}}_{0,\textsc{H}}^{(\phi)}(r;\omega)\,\tilde{G}_{0,\textsc{R}}^{(\phi)*}(r';\omega)\,\tilde{G}_{\textsc{R}}^{(Q)*}(\omega)\,e^{-i\omega(t-t')}\,,\label{E:keoruhs7}\\
	\frac{1}{2}\langle\bigl\{\hat{\phi}_{\text{h}}(x'),\hat{\phi}_{\textsc{r}}(x)\bigr\}\rangle_{0}&=e^{2}\int^{\infty}_{-\infty}\!\frac{d\omega}{2\pi}\;\prescript{\textsc{vac}}{}{\tilde{G}}_{0,\textsc{H}}^{(\phi)}(r';\omega)\,\tilde{G}_{0,\textsc{R}}^{(\phi)}(r;\omega)\,\tilde{G}_{\textsc{R}}^{(Q)}(\omega)\,e^{-i\omega(t-t')}\,,\label{E:keoruhs8}\\
	\frac{1}{2}\langle\bigl\{\hat{\phi}_{\textsc{r}}(x),\hat{\phi}_{\textsc{r}}(x')\bigr\}\rangle_{0}&=e^{4}\int^{\infty}_{-\infty}\!\frac{d\omega}{2\pi}\;\prescript{\textsc{vac}}{}{\tilde{G}}_{0,\textsc{H}}^{(\phi)}(0;\omega)\,\tilde{G}_{0,\textsc{R}}^{(\phi)}(r;\omega)\,\tilde{G}_{0,\textsc{R}}^{(\phi)*}(r';\omega)\,\tilde{G}_{\textsc{R}}^{(Q)}(\omega)\,\tilde{G}_{\textsc{R}}^{(Q)*}(\omega)\,e^{-i\omega(t-t')}\notag\\
	&=e^{2}\int^{\infty}_{-\infty}\!\frac{d\omega}{2\pi}\;\operatorname{sgn}(\omega)\,\operatorname{Im}\tilde{G}_{\textsc{R}}^{(Q)}(\omega)\,\tilde{G}_{0,\textsc{R}}^{(\phi)}(r;\omega)\,\tilde{G}_{0,\textsc{R}}^{(\phi)*}(r';\omega)\,e^{-i\omega(t-t')}\,,\label{E:keoruhs9}
\end{align}
at late times. It is of special importance that in \eqref{E:keoruhs9} we have used the fluctuation-dissipation relations~\cite{QTD1,QRad,NLFDR,CPR2D,CPR4D} of the vacuum fluctuations of the field $\hat{\phi}_{\text{h}}$ and the fluctuating internal degree of freedom $\hat{q}$ to rewrite part of its integrand as
\begin{equation}
	e^{2}\prescript{\textsc{vac}}{}{\tilde{G}}_{0,\textsc{H}}^{(\phi)}(0;\omega)\,\tilde{G}_{\textsc{R}}^{(Q)}(\omega)\,\tilde{G}_{\textsc{R}}^{(Q)*}(\omega)=\operatorname{sgn}(\omega)\,\operatorname{Im}\tilde{G}_{\textsc{R}}^{(Q)}(\omega)\,,
\end{equation}
which are essential to connect \eqref{E:keoruhs9}  with the combining contributions of \eqref{E:keoruhs7} and \eqref{E:keoruhs8}. These FDRs do not exist for the classical component $\varphi_{\text{h}}$ and $\mathsf{Q}$.

Putting this statement in a more explicit way, we have
\begin{align}
	\lim_{x'\to x}\frac{\partial^{2}}{\partial r\partial t'}\eqref{E:keoruhs7}+\eqref{E:keoruhs8}&=e^{2}\int^{\infty}_{-\infty}\!\frac{d\omega}{2\pi}\;\operatorname{sgn}(\omega)\,\omega^{2}\Bigl\{i\,\operatorname{Re}\tilde{G}_{0,\textsc{R}}^{(\phi)}(r;\omega)\,\tilde{G}_{0,\textsc{R}}^{(\phi)*}(r;\omega)\,\tilde{G}_{\textsc{R}}^{(Q)*}(\omega)\Bigr.\notag\\
	&\qquad\qquad\qquad\qquad\qquad\qquad-\Bigl.\operatorname{Im}\tilde{G}_{0,\textsc{R}}^{(\phi)}(r;\omega)\,\tilde{G}_{0,\textsc{R}}^{(\phi)}(r;\omega)\,\tilde{G}_{\textsc{R}}^{(Q)}(\omega)\Bigr\}\notag\\
	&=-i\,e^{2}\int^{\infty}_{-\infty}\!\frac{d\omega}{2\pi}\;\operatorname{sgn}(\omega)\,\omega^{2}\tilde{G}_{0,\textsc{R}}^{(\phi)*}(r;\omega)\,\tilde{G}_{0,\textsc{R}}^{(\phi)}(r;\omega)\,\tilde{G}_{\textsc{R}}^{(Q)}(\omega)\,,\label{E:dgkb1}
\end{align}
while
\begin{align}
	\lim_{x'\to x}\frac{\partial^{2}}{\partial r\partial t'}\eqref{E:keoruhs9}&=+i\,e^{2}\int^{\infty}_{-\infty}\!\frac{d\omega}{2\pi}\;\operatorname{sgn}(\omega)\,\omega^{2}\tilde{G}_{0,\textsc{R}}^{(\phi)*}(r;\omega)\,\tilde{G}_{0,\textsc{R}}^{(\phi)}(r;\omega)\,\tilde{G}_{\textsc{R}}^{(Q)}(\omega)\,,\label{E:dgkb2}
\end{align}
in the large $r$ limit. It is clearly seen that \eqref{E:dgkb1} and \eqref{E:dgkb2} cancel out. This in turn  implies for the fluctuating dynamics the net radiation power
\begin{equation}
	\frac{dw_{\textsc{rad}}}{d\tau}=0\,,\label{E:ruthgf}
\end{equation}
vanishes at late times over the spherical surface sufficiently far away from the atom. It is instructive to break down $dw_{\textsc{rad}}/d\tau$ into two quantum powers $P^{0}_{\times}$ and $P^{0}_{\textsc{r}}$, where
\begin{align}
	P^{0}_{\textsc{r}}(\infty)&=-\int\!d\Omega_{\bm{k}}\;r^{2}\,\eqref{E:dgkb2}=e^{2}\,\theta(t-r)\int^{\infty}_{-\infty}\!\frac{d\omega}{2\pi}\;\operatorname{sgn}(\omega)\,\omega^{2}\,\operatorname{Im}\tilde{G}_{\textsc{R}}^{(Q)}(\omega)\,\tilde{G}_{0,\textsc{R}}^{(\phi)*}(r;\omega)\,\tilde{G}_{0,\textsc{R}}^{(\phi)}(r;\omega)\notag\\
	&=e^{2}\,\theta(t-r)\int^{\infty}_{-\infty}\!\frac{d\omega}{2\pi}\;\operatorname{sgn}(\omega)\,\frac{\omega^{2}}{4\pi}\,\operatorname{Im}\tilde{G}_{\textsc{R}}^{(Q)}(\omega)\,,\\
	P^{0}_{\times}(\infty)&=-\int\!d\Omega_{\bm{k}}\;r^{2}\,\eqref{E:dgkb1}=-P^{0}_{\textsc{r}}(\infty)\,,
\end{align}
in the limit $t\gg\gamma^{-2}$. Here we have used the Fourier transforms of the Green's functions of the free field in the vacuum state,
\begin{align}
	\prescript{\textsc{vac}}{}{\tilde{G}}_{0,\textsc{H}}^{(\phi)}(\mathbf{r};\omega)&=\operatorname{sgn}(\omega)\,\operatorname{Im}\tilde{G}_{0,\textsc{R}}^{(\phi)}(r;\omega)\,,&\tilde{G}_{0,\textsc{R}}^{(\phi)}(\mathbf{r};\omega)&=\frac{e^{+i\omega r}}{4\pi r}\,,
\end{align}
and the parity properties of the Fourier transforms of the Green's functions. Eq.~\eqref{E:ruthgf} tells that the outgoing radiating power $P^{0}_{\textsc{r}}$ is equal to the incoming power $P_{\times}$, so that there is no net energy flow to spatial infinity.

The quantum power $P^{0}_{\textsc{r}}$ results purely from the radiation field $\hat{\phi}_{\textsc{r}}$, emitted by the atom when its internal dynamics is driven by the vacuum fluctuations of the free field $\hat{\phi}_{\text{h}}$. The other quantum power $P_{\times}$ has an intriguing nature. From \eqref{E:keoruhs7} and \eqref{E:keoruhs8}, we note it is the consequence of the correlation between $\hat{\phi}_{\textsc{r}}$ and $\hat{\phi}_{\text{h}}$. This correlation is propagated from the vacuum fluctuations of $\hat{\phi}_{\text{h}}$ to the fluctuating internal dynamics $\hat{q}$ of the atom, and finally to $\hat{\phi}_{\textsc{r}}$. The existence of the FDRs provides just the right amount of correlation to allow both powers to cancel out. This subtle cancellation at spatial infinity in the quantum degree of freedom of the field is necessary, even fascinating,  when we notice that there is a corresponding cancellation between the noise power $P_{\xi}^{0}(\infty)$ and the dissipation power $P_{\gamma}^{0}(\infty)$ in \eqref{E:dgnksdger} at the location of the atom, and that $P^{0}_{\times}(\infty)$ is equal to $P_{\xi}^{0}(\infty)$ in magnitude. This shows that when the  equilibrium state is reached, the steady incoming  energy flow at spatial infinity is  exactly what furnishes the energy power pumped into the atom. The balance of energy flow at spatial infinity and inside the atom is a powerful manifestation of the self-consistency condition locked inside the FDRs.

Comparing with the classical radiation powers, we take the product of the two equations  $\eqref{E:oeiufhd}$ and $\eqref{E:dkgsghfe}$,   and place  \eqref{E:dgkb1} side by side. We at once note that 1) the latter is time independent while the former is not even stationary in time, 2) the factors $\widetilde{G}_{0,\textsc{R}}^{(\phi)}(\mathbf{r};\omega)$ and $\widetilde{G}_{\textsc{R}}^{(Q)*}(\omega)\widetilde{G}_{0,\textsc{R}}^{(\phi)*}(r;\omega)$ for the classical power are not within the same integrand as in the  quantum power \eqref{E:dgkb1}, 3) there are no FDRs for the mean field $\varphi_{\text{h}}$ and the mean internal degree of freedom $\mathsf{Q}$ to relate $\eqref{E:dkgsghfe}$ to $\eqref{E:oeiufhd}$, and 4) since for this case the mean internal dynamics does not equilibrate, there is no definitive connection between $\mathcal{P}_{\xi}$ and $\mathcal{P}_{\gamma}$ as well as the corresponding radiation powers $\mathcal{P}_{\textsc{r}}$ and $\mathcal{P}_{\times}$; everything remains time-dependent.

In summary  we have shown that classical radiation and quantum radiation can be fused into a unified  formalism by the coherent state description. Depending on the late-time behavior of the free mean field, we have shown that if the free mean field is periodic, then the field radiated by the atom is essentially classical if the coherent parameter is much greater than unity. There will be nonvanishing classical radiation energy flowing outward to,  or inward from,  spatial infinity.  {The atom behaves similar to a dipole antenna}. Due to this nonstationary nature, the internal dynamics of the atom or the net radiation energy flow never comes to equilibrium. Here we would like to emphasize that our setup is a little different from the one used in~\cite{Jackson, Rohrlich}, where the {point} charge  {follows a prescribed external trajectory $\bm{z}(t)$ controlled} by an external agent,  not by the field that takes part in the interaction, so the contribution of $\mathcal{P}_{\times}$ is of no concern. On the other hand, if the free mean field diminishes at late times, then the radiation power is mainly of quantum nature. The existence of the fluctuation-dissipation relations for the quantum fluctuation dynamics enforces a subtle correlation between the free quantum field $\hat{\phi}_{\text{h}}$ and the radiation field $\hat{\phi}_{\textsc{r}}$. We thus find at late times there is no net quantum energy flow to spatial infinity. The radiated quantum power is delicately balanced by an  incoming quantum energy flow due to the aforementioned correlation. This shows very distinct late-time behavior in the radiated energy flow between the classical deterministic dynamics and the quantum fluctuation dynamics.

\section{conclusion}

As mentioned in the Introduction, this work continues what we began in Paper I,  extending  the study of quantum radiation from a stationary atom's internal degree of freedom to the appearance of radiation and  radiation reactions at both the quantum and the classical levels.   As the title of this paper suggests, our goal is to paint a continuum landscape starting from vacuum fluctuations in the quantum field to quantum dissipation in the atom resulting from its emittance of quantum radiation,  to classical radiation and classical radiation reaction.  Two major themes are presented: 
The first theme starts with placing the quantum field in a coherent state so we will have a unified formalism to simultaneously treat the classical and the quantum fields.  In the context of the atom-field interaction, one typically expects that when the coherent parameter of the field state is far greater than unity, both the field and the atom it interacts with  behave essentially like classical systems. This is a natural setting for our stated purpose and the results are largely as expected, except for a few caveats. For one, when one deals with the field initially in a multi-mode coherent state, it is true that the mean field looks like a classical field and its interaction with the internal degree of freedom of the atom will induces mean dynamics to it. For large coherent parameters, the atomic mean internal dynamics can dominate over the accompanying fluctuation dynamics. However, depending on  the frequency spectrum or the choice of the mode-dependent coherent parameters of the field, the mean field of physical interest to us, at late times, can be either periodic or decaying with time.  In the former case, the classical components of the field and the atomic internal dynamics will remain periodic and dwarf their quantum counterparts, so the atom's internal motion will not settle down to an equilibrium state. By contrast, in the latter case the atom will have a vanishing mean dynamics at late times,  where the corresponding fluctuating quantum evolution endures. Thus, the coherent state of the field will lead to quantum dynamics for the atom's internal degree of freedom, rather than the expected classical dynamics. Furthermore the internal dynamics will approach an equilibrium state, independent of the initial setting of the internal degree of freedom. 

{The second theme is subtler.  In the current configuration, the classical and the quantum components of the field and the atomic internal dynamics are fully decoupled. The mean (free) field drives the internal degree of freedom, here modeled by a harmonic oscillator. The {induced non-uniform internal motion} causes the atom to emit radiation, the reaction of which results in a  radiation reaction or self-force,  counteracting the motion of the atom's internal degree freedom. The classical  radiation will propagate to spatial infinity.   Up to this point, everything we have described is classical and deterministic, {just as in classical electrodynamics}.   On the other hand, we also have a quantum component in the internal motion of the atom induced by the vacuum fluctuations of the field in its coherent state. This component incites random motion in the atom's internal degree of freedom, which also causes the atom to radiate. Compared to the mean dynamics, the radiation in this case is purely quantum mechanical because the random motion is driven by the vacuum field fluctuations, and its associated reactive force (quantum radiation reaction or quantum self-force) is certainly of quantum nature. Thus from this delineation, we are able to make correct linkage between quantum (vacuum) fluctuation of the field and the corresponding quantum radiation reaction, whose origin is the stochastic motion driven by the field fluctuations.}

{A stronger connection can be made through examining the energy flux exchanges between the atom and the field. When the field fluctuations drive the atom's internal degree of freedom, an accompanying energy flows into the atom. The frictional force due to reaction of the emitted radiation on the other hand drains the atom's energy to the field surrounding the atom. The far field component of the radiated field will propagate away from the atom and in principle transport part of the atom's energy to spatial infinity. Since all the relevant physical quantities contain both the classical and the quantum components, the late-time behavior of the radiated classical far field at places sufficiently far away from the atom depends on the free mean field around the atom. When the free field is periodic, a spatially fixed probe far away from the atom will receive an energy flux periodic in time. The signal is typically much stronger than the noise level from the quantum component of the far field. On the other hand, if the mean field decreases with time, then this probe will receive a classical energy flux decaying with time, and in the end only {the quantum component of the radiation power will endure. Thus the probe may pick up  a weak energy flow exclusively associated with quantum radiation field fluctuations.}

The most drastic difference lie in the statistical characteristics of the classical and quantum components. The classical components are deterministic and at late times will be either periodic or falling off to zero; they do not equilibrate. The quantum components are stochastic in nature,  and for the presently studied configuration, they are correlated and will result in dynamical equilibration. Our results show that at late times the quantum component of the rate of energy pumped into the atom by the field is  balanced by the corresponding component of the rate of energy lost to the field.  The more intriguing and revealing fact is,  what we discovered in Paper I is repeated here, namely that,  at a point sufficiently far away from the atom, the energy carried away by the quantum component of the far field radiated from the atom is also compensated by another incoming energy flux related to the correlation between the radiated field far away from the atom, and the free field around the atom. This subtle correlation is a manifestation of two sets of fluctuation-dissipation relations; one associated with the free field and the other with the atom's internal degree of freedom. These relations govern the amount of correlation needed to balance the net radiated power to spatial infinity. Thus in fact a probe sitting at rest sufficiently far away will not see any energy flow to spatial infinity from a stationary atom. This in addition indicates that the contribution of the quantum field fluctuations is compensated by the counterpart from quantum dissipation (or quantum radiation reaction), the reactive force of quantum radiation. Such cancellation in general is not available for the classical radiation from a stationary atom because no FDR exists  for the mean internal dynamics of the atom and the mean dynamics of the field. }\\\\

\noindent {\bf Acknowledgments} J.-T. Hsiang is supported by the Ministry of Science and Technology of Taiwan, R.O.C. under Grant No.~MOST 110-2811-M-008-522. 

\appendix
\section{coherent state in a nutshell}\label{S:rtghjvdf}
The coherent state $\lvert\alpha\rangle$ is the eigenstate of the annihilation operator $\hat{a}$ with the complex eigenvalue $\alpha$,
\begin{equation}
	\hat{a}\,\lvert\alpha\rangle=\alpha\,\lvert\alpha\rangle\,,
\end{equation}
which implies that an expansion in terms of the number states
\begin{equation}
	\lvert\alpha\rangle=e^{-\frac{\lvert\alpha\rvert^{2}}{2}}\sum_{n=0}^{\infty}\frac{\alpha^{n}}{\sqrt{n!}}\,\lvert n\rangle\,,
\end{equation}
with $\hat{N}\,\lvert n\rangle=n\,\lvert n\rangle$ and $\hat{N}=\hat{a}^{\dagger}\hat{a}$. It is then straightforward to see that the coherent state is not orthogonal because for any two coherent states $\lvert\alpha\rangle$ and $\lvert\beta\rangle$, they has nonzero overlap,
\begin{align}
	\langle\beta\vert\alpha\rangle=\exp\biggl\{-\frac{\lvert\alpha\rvert^{2}}{2}-\frac{\lvert\beta\rvert^{2}}{2}+\beta^{*}\alpha\biggr\}\,.
\end{align}
However the coherent states tend to become approximately orthogonal for values of $\alpha$ and $\beta$ are sufficiently different. Thus in general the coherent state is not linearly independent from one another. The coherent state forms a complete set in the sense that
\begin{equation}
	\int\!\frac{d^{2}\alpha}{\pi}\;\lvert\alpha\rangle\langle\alpha\rvert=1\,.
\end{equation}

Alternatively, we may define the coherent state in terms of the \textit{unitary} operator $\hat{D}(\alpha)$ 
\begin{equation}
	\hat{D}(\alpha)=\exp\bigl(\alpha\,\hat{a}^{\dagger}-\alpha^{*}\,\hat{a}\bigr)\,.
\end{equation}
such that $\lvert\alpha\rangle=\hat{D}(\alpha)\,\lvert0\rangle$. It can be readily verified that 
\begin{align}
	\hat{a}\,\lvert\alpha\rangle=\alpha\,\lvert\alpha\rangle\,,
\end{align}
indeed, is the eigenstate of $\hat{a}$.

By the BCH formula
\begin{equation}\label{E:rbgdhfjg}
	e^{\hat{X}}\,\hat{Y}\,e^{-\hat{X}}=\hat{Y}+\bigl[\hat{X},\hat{Y}\bigr]+\frac{1}{2!}\,\bigl[\hat{X},\bigl[\hat{X},\hat{Y}\bigr]\bigr]+\frac{1}{3!}\,\bigl[\hat{X},\bigl[\hat{X},\bigl[\hat{X},\hat{Y}\bigr]\bigr]\bigr]+\cdots\,,
\end{equation}
we find 
\begin{align}
	\hat{D}(\alpha)\,\hat{a}\,\hat{D}^{-1}(\alpha)&=\hat{a}-\alpha\,,
\end{align}
such that
\begin{align}
	\langle\hat{a}^{2}\rangle_{\alpha}&=\alpha^{2}\,,&\langle\hat{a}^{\dagger2}\rangle_{\alpha}&=\alpha^{*2}\,,&\langle\hat{a}^{\dagger}\hat{a}\rangle_{\alpha}&=\lvert\alpha\rvert^{2}\,,&\langle\hat{a}\hat{a}^{\dagger}\rangle_{\alpha}&=\lvert\alpha\rvert^{2}+1\,,
\end{align}
where $\langle\cdots\rangle_{\alpha}$ is understood as $\langle\alpha\vert\cdots\vert\alpha\rangle$.

In the context of oscillator dynamics, the Heisenberg equation of motion of the harmonic oscillator is given by
\begin{equation}\label{E:dgbksur}
	\ddot{\hat{Q}}(t)+\omega^{2}\hat{Q}(t)=0\,,
\end{equation}
where $\omega$ is the oscillating frequency, so its expectation value takes the form
\begin{equation}
	\ddot{\mathsf{Q}}(t)+\omega^{2}\,\mathsf{Q}(t)=0\,,\label{E:fkgbseur}
\end{equation}
with $\mathsf{Q}(t)=\langle\hat{Q}(t)\rangle_{\alpha}$. From \eqref{E:dgbksur}, the displacement operator $\hat{Q}$ and the conjugate momentum $\hat{P}$ of the quantum harmonic oscillator evolves according to
\begin{align}
	\hat{Q}(t)&=\sqrt{\frac{\hbar}{2m\omega}}\,\bigl(\hat{a}^{\dagger}\,e^{+i\omega t}+\hat{a}\,e^{-i\omega t}\bigr)\,,&\hat{P}(t)=i\sqrt{\frac{\hbar m\omega}{2}}\,\bigl(\hat{a}^{\dagger}\,e^{+i\omega t}-\hat{a}\,e^{-i\omega t}\bigr)\,,
\end{align}
so we immediately have
\begin{align}\label{E:roeutjgskf}
	\mathsf{Q}(t)&=\sqrt{\frac{\hbar}{2m\omega}}\,\bigl[\alpha^{*}\,e^{+i\omega t}+\alpha\,e^{-i\omega t}\bigr]\,,&\mathsf{P}(t)&=i\sqrt{\frac{\hbar m\omega}{2}}\,\bigl(\alpha^{*}\,e^{+i\omega t}-\alpha\,e^{-i\omega t}\bigr)\,.
\end{align}
Thus, Eqs.~\eqref{E:fkgbseur} and \eqref{E:roeutjgskf} implies the mean dynamics of the quantum harmonic oscillator in the coherent state behaves like a classical harmonic oscillator, with its amplitude proportional to $\alpha$.

On the other hand, the quantum harmonic oscillator in the coherent state has nonzero dispersions for its canonical variables,
\begin{align}
	\langle\Delta\hat{Q}^{2}\rangle_{\alpha}&=\frac{\hbar}{2m\omega}\,,&\langle\Delta\hat{P}^{2}\rangle_{\alpha}&=\frac{\hbar m\omega}{2}\,,&\frac{1}{2}\,\langle\bigl\{\Delta\hat{Q},\Delta\hat{P}\bigr\}\rangle_{\alpha}&=0\,.
\end{align}
We right away conclude
\begin{align}
	\langle\Delta\hat{H}\rangle_{\alpha}&=\frac{\hbar\omega}{2}\,,&\langle\Delta\hat{Q}^{2}\rangle_{\alpha}\langle\Delta\hat{P}^{2}\rangle_{\alpha}-\Bigl[\frac{1}{2}\,\langle\bigl\{\Delta\hat{Q},\Delta\hat{P}\bigr\}\rangle_{\alpha}\Bigr]^{2}&=\frac{\hbar^{2}}{4}\,.
\end{align}
Hence although its mean dynamics follows the trajectory of a classical harmonic oscillator, the quantum harmonic oscillator in the coherent state has the minimal uncertainty exclusively due to the zero-point quantum fluctuations. That is why the coherent state is sometimes viewed as the most classical quantum state, in particular, when $\lvert\alpha\rvert\gg1$.

For a given coherent state $\lvert\alpha\rangle$ of the oscillator, the probability of finding the oscillator in the $n^{\text{th}}$ excited state is 
\begin{equation}\label{E:kgfjbdfe}
	\mathfrak{P}(n)=\lvert\langle n\vert\alpha\rangle\rvert^{2}=\frac{\lvert\alpha\rvert^{2n}}{n!}\,e^{-\lvert\alpha\rvert^{2}}\,.
\end{equation}
If we let $N$ be the average number of excitations
\begin{equation}
	N=\langle\hat{N}\rangle_{\alpha}=\langle\hat{a}^{\dagger}\hat{a}\rangle_{\alpha}=\lvert\alpha\rvert^{2}\,,
\end{equation}
then we can write the probability \eqref{E:kgfjbdfe} in terms of $N$,
\begin{equation}
	\mathfrak{P}(n)=\frac{N^{n}}{n!}\,e^{-N}\,.
\end{equation}
This corresponds to the Poisson distribution. This implies that
\begin{equation}
	\frac{\langle\Delta\hat{N}^{2}\rangle_{\alpha}^{\frac{1}{2}}}{\langle\hat{N}\rangle_{\alpha}}=\frac{1}{\lvert\alpha\rvert}\,.
\end{equation}

If we define
\begin{align}
	\hat{Q}^{(-)}(t)&=\sqrt{\frac{\hbar}{2m\omega}}\,\hat{a}\,e^{-i\omega t}\,,&&\text{and}&\hat{Q}^{(+)}(t)&=\sqrt{\frac{\hbar}{2m\omega}}\,\hat{a}^{\dagger}\,e^{+i\omega t}\,,
\end{align}
then we may introduce the temporal degree of second-order coherence by
\begin{equation}\label{E:bnfsgo}
	g^{(2)}(\tau)=\frac{\langle\hat{Q}^{(+)}(t)\hat{Q}^{(+)}(t+\tau)\hat{Q}^{(-)}(t+\tau)\hat{Q}^{(-)}(t)\rangle}{\langle\lvert\hat{Q}^{(-)}(t)\rvert^{2}\rangle\langle\lvert\hat{Q}^{(-)}(t+\tau)\rvert^{2}\rangle}\,.
\end{equation}
for a quantum state in which the expectation values in \eqref{E:bnfsgo} are taken. For a coherent state, we find $g^{(2)}(\tau)=1$ for all $\tau$. In comparison, the classical light has $g^{(2)}(\tau)\leq g^{(2)}(0)$ but $g^{(2)}(0)\leq1$. In fact, the coherent state owns all orders of coherence because all of the degrees of the higher-order coherence for the coherent state are equal to unity, where the degree of the $n^{\text{th}}$-oder coherence is defined by
\begin{align}
	g^{(2)}(t_{1},\cdots,t_{n})=\frac{\langle\hat{Q}^{(+)}(t_{1})\cdots\hat{Q}^{(+)}(t_{n})\hat{Q}^{(-)}(t_{n})\cdots\hat{Q}^{(-)}(t_{1})\rangle}{\langle\lvert\hat{Q}^{(-)}(t_{1})\rvert^{2}\rangle\cdots\langle\lvert\hat{Q}^{(-)}(t_{n})\rvert^{2}\rangle}\,.
\end{align}

For the multi-mode coherent state, suppose we have a Hermitian operator $\hat{O}(t)$ expanded in terms of the complex mode functions $u_{\bm{k}}(t)$, labelled by $\bm{k}$, and the associated time independent annihilation and creation operators $\hat{a}_{\bm{k}}^{\vphantom{\dagger}}$, $\hat{a}_{\bm{k}}^{\dagger}$,
\begin{equation}
	\hat{O}(t)=\sum_{\bm{k}}\hat{a}_{\bm{k}}^{\vphantom{\dagger}}\,u_{\bm{k}}^{\vphantom{*}}(t)+\hat{a}_{\bm{k}}^{\dagger}\,u_{\bm{k}}^{*}(t)\,.
\end{equation}
For the coherent state $\lvert\{\alpha\}\rangle$, abbreviated for $\lvert\{\alpha\}\rangle=\lvert\alpha_{\bm{k}_{1}}\rangle\otimes\lvert\alpha_{\bm{k}_{2}}\rangle\otimes\cdots$, we immediately have
\begin{equation}
	\mathsf{O}(t)=\langle\{\alpha\}\vert\,\hat{O}\,\vert\{\alpha\}\rangle=\sum_{\bm{k}}\alpha_{\bm{k}}^{\vphantom{\dagger}}\,u_{\bm{k}}^{\vphantom{*}}(t)+\alpha_{\bm{k}}^{\dagger}\,u_{\bm{k}}^{*}(t)\,.
\end{equation}
The corresponding Hadamard function $G_{\textsc{H}}^{(O)}(t,t')$ is
\begin{align}
	G_{\textsc{H}}^{(O)}(t,t')=\frac{1}{2}\,\langle\{\alpha\}\vert\bigl\{\hat{O}(t), \hat{O}(t')\bigr\}\vert\{\alpha\}\rangle&=\sum_{\bm{k},\bm{k}'}\bigl[\alpha_{\bm{k}}^{\vphantom{\dagger}}\,u_{\bm{k}}^{\vphantom{*}}(t)+\alpha_{\bm{k}}^{\dagger}\,u_{\bm{k}}^{*}(t)\bigr]\bigl[\alpha_{\bm{k}'}^{\vphantom{\dagger}}\,u_{\bm{k}'}^{\vphantom{*}}(t')+\alpha_{\bm{k}'}^{\dagger}\,u_{\bm{k}'}^{*}(t')\bigr]\notag\\
	&\qquad\qquad\qquad\qquad+\frac{1}{2}\sum_{\bm{k}}\bigl[u_{\bm{k}}^{\vphantom{*}}(t)u_{\bm{k}'}^{*}(t')+u_{\bm{k}}^{*}(t)u_{\bm{k}'}^{\vphantom{*}}(t')\bigr]\notag\\
	&=\mathsf{O}(t)\mathsf{O}(t')+\frac{1}{2}\,\langle0\vert\bigl\{\hat{O}(t), \hat{O}(t')\bigr\}\vert0\rangle\,,
\end{align}
where we have used
\begin{align}
	\langle\{\alpha\}\vert\,\hat{a}_{\bm{k}}\hat{a}_{\bm{k}'}\,\vert\{\alpha\}\rangle&=\alpha_{\bm{k}}\alpha_{\bm{k}'}\,,\\
	\langle\{\alpha\}\vert\,\hat{a}_{\bm{k}}^{\vphantom{\dagger}}\hat{a}_{\bm{k}'}^{\dagger}\,\vert\{\alpha\}\rangle&=\langle\{\alpha\}\vert\,\hat{a}_{\bm{k}'}^{\dagger}\hat{a}_{\bm{k}}^{\vphantom{\dagger}}+\delta_{\bm{k}\bm{k'}}\,\vert\{\alpha\}\rangle=\alpha_{\bm{k}}^{\vphantom{*}}\alpha_{\bm{k}'}^{*}+\delta_{\bm{k}\bm{k'}}\,.
\end{align}



\newpage
\begin{thebibliography}{999}
	
\bibitem{QRad}
	J.-T. Hsiang, and B. L. Hu, \textit{Atom-field interaction: From vacuum fluctuations to quantum radiation and quantum dissipation or radiation reaction}, Physics {\bf1}, 430 (2019).
		
\bibitem{MilBook} 
	P. W. Milonni, {\sl The Quantum Vacuum: An Introduction to Quantum Electrodynamics} (Academic Press, 1993).

\bibitem{PassanteBook} 
	G. Compagno, R. Passante, and F. Persico, {\sl Atom-Field Interactions and Dressed Atoms} (Cambridge Press, Cambridge, 1995).

\bibitem{CTBook} 
	C. Cohen-Tannodji, B. Diu, and F. Laloe, {\sl Quantum Mechanics} (Wiley, New York, 1991).

\bibitem{ScullyBook} 
	M. O. Scully, and M. S. Zubairy, {\sl Quantum Optics} (Cambridge Press, Cambridge, 1998).
		
\bibitem{RHA} 
	A. Raval, B. L. Hu, and J. Anglin, {\it Stochastic theory of accelerated detectors in a quantum field}, Phys. Rev. D \textbf{53}, 7003 (1996).

\bibitem{RHK} 
	A. Raval, B. L. Hu, and D. Koks, {\it Near-thermal radiation in detectors, mirrors, and black holes: A stochastic approach}, Phys. Rev. D \textbf{55}, 4795 (1997).

\bibitem{JH1} 
	P. R. Johnson, and B. L. Hu, {\it Stochastic theory of relativistic particles moving in a quantum field: Scalar Abraham-Lorentz-Dirac-Langevin equation, radiation reaction, and vacuum fluctuations}, Phys. Rev. D \textbf{65}, 065015  (2002).

\bibitem{CPR2D}
	J.-T. Hsiang, B. L. Hu, and S.-Y. Lin, {\it Fluctuation-dissipation and correlation-propagation relations from the nonequilibrium dynamics of detector-quantum field systems}, Phys. Rev. D \textbf{100}, 025019 (2019). 

\bibitem{CPR4D}  
	J.-T. Hsiang, B. L. Hu, S.-Y. Lin, and K. Yamamoto, {\it Fluctuation-dissipation and correlation-propagation relations in (1+3)D moving detector-quantum field systems}, Phys. Lett. B \textbf{795}, 694 (2019).
		
\bibitem{Ja74}		
	R. Jackiw, \textit{Functional evaluation of the effective potential}, Phys. Rev. D {\bf9}, 1686  (1974).

\bibitem{NLFDR}
		J.-T. Hsiang, and B. L. Hu, \textit{Fluctuation-dissipation relation from the nonequilibrium dynamics of a nonlinear open quantum system}, Phys. Rev. D {\bf101}, 125003 (2020).
		

\bibitem{FDRSq}
	J.-T. Hsiang, and B. L. Hu, \textit{Fluctuation–dissipation relation for a quantum Brownian oscillator in a parametrically squeezed thermal field}, Ann. Phys. {\bf433}, 168594 (2021).

\bibitem{QTD1}
	 J.-T. Hsiang, C. H. Chou, Y. Suba\c{s}\i, and B. L. Hu, \textit{Quantum thermodynamics from the nonequilibrium dynamics of open systems: Energy, heat capacity, and the third law}, Phys. Rev. E {\bf97}, 012135 (2018).

\bibitem {Unr76}
	W. G. Unruh, {\it Notes on black-hole evaporation}, Phys. Rev. {\bf D 14}, 870 (1976).

\bibitem{DeW79} 
	B. S. DeWitt, in {\sl General Relativity: an Einstein Centenary Survey}, edited by S. W. Hawking and W. Israel (Cambridge Press, Cambridge, 1979).

\bibitem{Fulling19} 
	A. G. S. Landulfo,  S. A.  Fulling,  and G. E. A. Matsas, {\it Classical and quantum aspects of the radiation emitted by a uniformly accelerated charge: Larmor-Unruh reconciliation and zero-frequency Rindler modes},  Phys. Rev. D {\bf100}, 045020  (2019).

\bibitem{LH06}
	S.-Y. Lin, and B. L. Hu, \textit{Accelerated detector–quantum field correlations: From vacuum fluctuations to radiation
flux}, Phys. Rev. D  {\bf73}, 124018 (2006). 

\bibitem{Jackson}
	J. D. Jackson, \textsl{Classical Electrodynamics, 2nd Ed.} (Wiley, New York, 1975).

\bibitem{Rohrlich}
	F. Rohrlich, \textsl{Classical Charged Particles - Foundation of their Theories} (Westview, Colorado, 1990).
		

\end{thebibliography}
\end{document}